\documentclass[11pt,showpacs]{revtex4}
\input epsf.sty

\def\comment#1{}
\def\aut#1{#1}
\def\yy{\zeta}
\newcommand{\sfrac}[2]{\raisebox{0.095ex}{\scriptsize${\frac{#1}{#2}}$}}
\newcommand{\sbf}[1]{\mbox{{\scriptsize$\bf{#1}$}}}
\def\mn#1{*\marginpar{*\tiny{#1}}}
\def\mn#1{}
\def\E{{\mathcal E}}
\def\Ds{D_s}

\def\lfrac#1#2{{#1/#2}}
\usepackage{graphicx}
\begin{document}

\title{Electron-Positron Pair Production in Space- or Time-Dependent Electric
Fields}
\author{Hagen Kleinert$^{(a,b)}$, Remo Ruffini$^{(b)}$ and She-Sheng Xue$^{(b)}$}
\affiliation{$^{(a)}$Institut f{\"u}r Theoretische Physik, Freie Universit\"at Berlin, 14195 Berlin, Germany}
\affiliation{$^{(b)}$ICRANeT Piazzale della Repubblica, 10 -65122, Pescara, and\\
Dipartimento di Fisica, University of Rome ``La Sapienza", P.le A. Moro 5, 00185 Rome, Italy}


\date{Received version \today}

\begin{abstract}
Treating the production of electron and positron pairs
by a strong electric
field from the vacuum
as
a quantum tunneling process
we derive,
in semiclassical
approximation,
a
general  expression for the
pair production rate in a $z$-dependent
electric field $E(z)$ pointing in
the $z$-direction.
We also allow for a smoothly varying magnetic field
parallel to $E(z)$.
The result is applied to a
confined field $E(z)\not=0$ for $|z|\lesssim \ell $,
a semi-confined field $E(z)\not=0$ for $ z\gtrsim 0 $, and
a linearly increasing field $E(z)\sim z$. The
boundary effects of
the confined fields on
pair-production rates are exhibited.
A simple variable change in all formulas
leads to results for electric fields
depending on time rather than space.

{In addition, we discuss
tunneling processes
in which
empty atomic bound states
are spontaneously filled
by negative-energy electrons
from the vacuum
under positron emission.
In particular,
we calculate the rate at which the
atomic levels
of a bare nucleus of finite size $r_{\rm n}$ and large $Z\gg 1$
are filled
by spontaneous pair creation. }
\end{abstract}
\maketitle

\section{Introduction}\label{intro}

The creation of electron-positron pairs
 from the vacuum by an external uniform
electric field in spacetime was
first studied by Sauter \cite{sauter}
as a quantum tunneling process.
Heisenberg and Euler~\cite{euler} extended his result
by calculating an effective Lagrangian
from the
Dirac theory for electrons in
a constant electromagnetic field. A more elegant
quantum field theoretic reformulation was given by Schwinger \cite{schwinger} based on calculations of
who
calculated
within
{\it Quantum Electrodynamics} (QED)
the
 one-loop effective action in a constant electromagnetic field.
 A detailed review and relevant
references can be found in Refs.~\cite{dunne} and \cite{rvx2007}.

Apart from its purely theoretic interest,
 the pair-production in nonuniform fields is
experimentally relevant
in collisions of laser beams \cite{r} and heavy ions \cite{z,gbook},
and as a possible
explanation
 of the powerful Gamma Ray Bursts in
astrophysics \cite{grb,rvx2007}. It is also important for understanding the plasma oscillations \cite{osci}
of electrons and positrons after their  creation in electric fields.

The rate of pair-production may be split into
an exponential and a pre-exponential factor.
The exponent is determined by the
 classical trajectory of the tunneling particle
in imaginary time which has the smallest action.
It plays the
same role as the activation energy
in a
Boltzmann factor
with
a ``temperature"
$\hbar $.
The
pre-exponential factor is determined
by the quantum fluctuations of the path around that trajectory.
At the
 semiclassical level, the latter is obtained
from the functional determinant
of the quadratic fluctuations.
It
can be calculated in closed form only for
a few classical paths \cite{hkpath}.
An efficient
technique for
doing this
is based on
the WKB wave functions, another on
solving
 the
Heisenberg
equations of motion
for the position operator in the external field \cite{hkpath}.

Given the difficulties
in calculating the pre-exponential factor,
only
a few
 nonuniform electric fields
in space or in time
have led to
analytic results
for the pair-production rate: (i) the electric field in the ${z}$-direction is confined in the space $x<x_0$, i.e.,
${\bf E}=E(x)\hat{\bf z}$ where $E(x)=E_0\theta(x_0-x)$
\cite{mv1988}; (ii) the electric field in the ${z}$-direction depends only on the light-cone coordinate $z_+=(t+z)/\sqrt{2}$,
i.e., ${\bf E}=E(z_+)\hat{\bf z}$ \cite{ttw2000}.
If the nonuniform field
has the form
$E(z)=E_0/\cosh^2(z)$, which we shall refer to as
a Sauter field,
the rate was calculated by
solving the Dirac equation \cite{nanozhni1970}
in the same way as
Heisenberg and Euler
did
for the constant electric field.
For general
space and time dependences,
only the exponential
factor
can be written down easily ---
the
fluctuation factor
is usually hard to calculate \cite{z}. In the
Coulomb field of heavy nucleus whose size is finite and charge $Z$ is
supercritical, the problem becomes even more difficult for bound states
being involved in pair production, and a lot of effort has
been spent
on this issue \cite{z,gbook,rfk78}. 

If the electric field has only a time dependence $E=E(t)$,
 both exponential and pre-exponential factors were approximately
computed by
Brezin and Itzykson using WKB methods
for the purely periodic field  $E(t)=E_0\cos\omega_0 t$ \cite{r20}.
The result was generalized
by Popov in Ref.~\cite{popov1972}
to more general time-dependent
fields $E(t)$.
After this, several time-independent
but space-dependent
fields were treated,
for instance
an electric field
between  two conducting plates \cite{ww1988},
and an electric field around a Reissner-Nordstr\"om black hole \cite{k2000}.

The semicalssical expansion was carried beyond the WKB
approximation
by calculating higher-order corrections in powers
of
$\hbar$
in Refs. \cite{kp2006} and \cite{kp2007}.
Unfortunately, these terms do not comprise all
corrections of the same orders
$\hbar$
as will be
explained at the end of Section \ref{@RATE}.

An alternative approach to the same problems
was recently proposed
by using
the worldline
formalism
\cite{s2001}, sometimes called the ``string-inspired formalism''.
This
formalism is closely related
to Schwinger's quantum field theoretic treatment of the tunneling problem, where
the
evaluation
of a fluctuation determinant
is required
involving
the {\it fields\/} of the particle
pairs created from the vacuum.
The worldline approach is special
technique for calculating precisely this
functional determinant.
Within the worldline
 formalism,
Dunne and Schubert \cite{ds2005}
calculated the exponential
factor and Dunne et al. \cite{dwgs2006}
gave the associated
prefactor
for various  field configurations: for instance
a spatially uniform, and
single-pulse field with a temporal Sauter shape
$\propto 1/\cosh^2  \omega  t$.
For general $z$-dependences,
a numerical calculation scheme
was proposed in Ref. \cite{gl2001}
and applied further in \cite{gk2005}.
For a multidimensional extension
of the techniques see
Ref. \cite{dw2006}.


In this article
 we derive a general expression
for the pair-production rate in nonuniform electric fields $E(z)$
pointing in the $z$-direction and varying
only along this direction. A simple variable change in all formulas
leads to results for electric fields
depending on time rather than space.
As examples, we
shall treat
three cases:
(i) a nonzero electric field confined to a region of size $\ell $,
i.e., $E(z)\not=0, |z|\lesssim \ell $ (Sauter field see Eq.~(\ref{sauterv}));
(ii) a nonzero electric field in a half-space, i.e.,
$E(z)\not=0, z\gtrsim 0 $ (see Eq.~(\ref{hv})); (iii)
an electric field
increasing linearly like
 $E(z)\sim z$. 
In addition
we shall study the process
of
negative-energy electrons tunneling
 into the bound states of an electric potential
with the emission of positrons. We consider two cases: (1)
the electric field $E(z)\sim z$ of harmonic potential $V(z)\sim z^2$; (2) the
radial Coulomb field $E(r)=eZ/r^2$
with large $Z$
outside
the nuclear radius $r_{\rm n}$.

\section{Semi-classical description of pair production}\label{semi}

\begin{figure}[th]
\begin{center}
\begin{picture}(105.64,184.645)
\def\fsz{\footnotesize}
\def\ssz{\scriptsize}
\def\tsz{\tiny}
\def\dst{\displaystyle}\unitlength1mm
\put(-20,0){\includegraphics[width=8cm,clip]{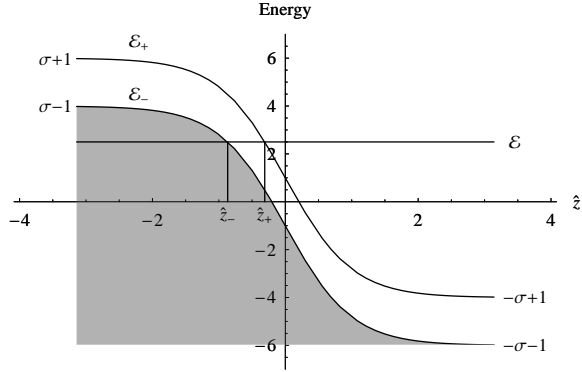}}
\end{picture}
\end{center}
\caption{
Positive- and negative-energy
spectra ${\mathcal E}_\pm(z)$
of Eq.~(\ref{energyl+-})
in units of $m_ec^2$,
with  $p_z=p_\perp =0$
as a function of $\hat z=z/\ell $
for the Sauter potential $V_\pm(z)$ (\ref{sauterv}) for $\sigma=5$.
}%
\label{sauterf}%
\end{figure}

\comment{
Since the pioneering work
of Sauter \cite{sauter},
Heisenberg and Euler \cite{euler},
and its elegant
quantum field theoretic reformulation by Schwinger \cite{schwinger},
}
The phenomenon of pair production in an external electric field
can be understood
as
a quantum-mechanical tunneling process of Dirac electrons \cite{Dir30,Dir33}.
In the original Dirac picture,
the electric field
bends the positive and negative-energy levels of the
Hamiltonian,
leading to a level-crossing
and a tunneling of the electrons
in the negative-energy band to the positive-energy band.
Let the field vector ${\bf E}(z)$
point in the ${z}$-direction.
In the one-dimensional potential energy
\begin{equation}
V(z)=-eA_0(z)
=e\int^z dz'E(z')
\label{@VEQ}\end{equation}
of an electron of charge $-e$,
the
classical
 positive and negative-energy spectra are
\begin{equation}
{\mathcal E}_\pm(p_z,{p}_\perp;z)=\pm\sqrt{(cp_z)^2+c^2{ p}_\perp^2+(m_ec^2)^2}+V(z),
\label{energyl+-}
\end{equation}
where $p_z$ is the momentum in
the ${z}$-direction, ${\bf p}_\perp$ the momentum
orthogonal to it, and $p_\perp\equiv |{\bf p}_\perp|$.
For a given energy
${\mathcal E}$, the tunneling takes place
from $z_-$ to $z_+$ determined by $p_z=0$ in Eq.~(\ref{energyl+-})
\begin{equation}
{\mathcal E}={\mathcal E}_+(0,{p}_\perp;z_+)=
{\mathcal E}_-(0,{p}_\perp;z_-).
\label{crossing}
\end{equation}
The
points
$z_\pm$
are the {\em turning points\/}
of the classical trajectories
 crossing from the positive-energy band
to the negative one at energy $\E$.
They satisfy
the equations
\begin{equation}
V(z_\pm)=\mp\sqrt{c^2p_\perp^2+m_e^2c^4}+{\mathcal E} .
\label{crosspoint+-}
\end{equation}
This energy-level crossing $\E$ is shown in Fig.~\ref{sauterf} for the Sauter potential
$V(z)\propto \tanh (z/\ell)$.

\subsection{WKB transmission probability for Klein-Gordon Field}

The probability of quantum tunneling  in the $z$-direction
is most easily studied
for a
scalar field which satisfies
the Klein-Gordon equation
\begin{eqnarray}
\left\{ \left[ i\hbar \partial _\mu+\frac{e}{c}A_\mu(z)\right] ^2
-m_e^2c^2
\right\}
\phi(x)=0,
\label{@KG0}\end{eqnarray}
where $x_0\equiv ct$. If there is only an electric field in the $z$-direction
which varies only along $z$, we can
choose a vector potential with the only nonzero component
(\ref{@VEQ}),
and
make the ansatz
$\phi(x)=e^{-i{\cal E}t/\hbar }e^{i{\sbf p}_\perp {\sbf x}_\perp/\hbar } \phi_{{\sbf p}_\perp,\E}(z)$,
with a fixed momentum
${\bf p}_\perp$
in the $x,y$ direction and an energy ${\cal E}$, and Eq.~(\ref{@KG0})
becomes simply
\begin{eqnarray}
\left[-\hbar ^2\frac{d^2}{dz^2}+p_\perp^2
+m_e^2c^2-\frac{1}{c^2} \left[\E-V(z) \right]^2\right]  \phi_{{\sbf p}_\perp,\E}(z)=0.
\label{KG}\end{eqnarray}
By expressing the wave function
$ \phi_{{\sbf p}_\perp,\E}(z)$ as an exponential
\begin{equation}
 \phi_{{\sbf p
}_\perp,\E}(z)
=   {\mathcal C}\,e^{iS_{{\sbf p}_\perp,\E}/\hbar },
\end{equation}
where ${\mathcal C}$ is some normalization  constant,
the wave equation becomes a
Riccatti equation
for $S_{{\sbf p}_\perp,\E}$:
\begin{equation} \label{n4.5}
-i\hbar \partial _z^2
S_{{\sbf p}_\perp,\E}(z)+
[\partial _z
S_{{\sbf p}_\perp,\E}(z)]^2-p_z^2(z)=0.
\end{equation}
where the function $p_z(z)$ is the solution of
the equation
\begin{eqnarray}
p_z^2(z)=\frac{1}{c^2}\left[\E-V(z) \right]^2-p_\perp^2-m_e^2c^2.
\label{WKBr2}
\end{eqnarray}
The solution
of Eq. (\ref{n4.5})  can be found iteratively as an expansion in powers of $\hbar$:
\begin{equation} \label{4.13}
S_{{\sbf p}_\perp,\E}(z)=
S^{(0)}_{{\sbf p}_\perp,\E}(z)-i\hbar
S^{(1)}_{{\sbf p}_\perp,\E}(z)+(-i\hbar )^2
S^{(2)}_{{\sbf p}_\perp,\E}(z)+\dots~.
\end{equation}
Neglecting the expansion terms after $
S^{(1)}_{{\sbf p}_\perp,\E}(z)=-\log p^{1/2}_z(z) $ leads to the WKB approximation
for the wave functions of positive and negative energies can (see e.g.
\cite{LanLif75,PI})
\begin{eqnarray}
\phi^{\rm WKB}_{{\sbf p}_\perp,\E}(z)=
\frac{\mathcal C}{p^{1/2}_z(z)}e^{iS^{(0)}_{{\sbf p}_\perp,\E}(z)/\hbar }.
\label{WKBs}
\end{eqnarray}
 where  $S^{(0)}_{{\sbf p}_\perp,\E}(z)$
is the eikonal
\begin{eqnarray}
S^{(0)}_{{\sbf p}_\perp,\E}(z)=\int^zp_z(z')dz'.
\label{WKBs1}
\end{eqnarray}
Between the turning points
$z_-<z<z_+$,
whose positions are illustrated in
Fig.~\ref{sauterf},
the momentum
$p_z(z)$ is imaginary
and is useful to define the positive function
\begin{eqnarray}
 \kappa _z(z)\equiv  \sqrt{
p_\perp^2+m_e^2c^2-  \frac{1}{c^2}
\left[\E-V(z) \right]^2
}\geq 0.
\end{eqnarray}
The tunneling wave function in this regime is the linear combination
\begin{eqnarray}
\frac{{\mathcal C}}{2(\kappa_z)^{1/2}}\exp\left[ -\frac{1}{\hbar}\int^{z}_{z_-}\kappa_zdz\right]+
\frac{\bar {\mathcal C}}{2(\kappa_z)^{1/2}}\exp\left[ +\frac{1}{\hbar}\int^{z}_{z_-}\kappa_zdz\right].
\label{WKBt}
\end{eqnarray}
Outside
the turning points, i.e., for
$z<z_-$ and
$z>z_+$,
there exist negative-energy and  positive-energy solutions for
$\E<\E_-$ and
$\E>\E_+$
for positive $p_z$.
On the left-hand side
of $z_-$,
the general solution is
a linear combination
of an incoming wave  running to the right and outgoing wave running to the left:
\begin{eqnarray}
\frac{\mathcal C_+}{(p_z)^{1/2}}\exp
\left[
 \frac{i}{\hbar}\int^zp_zdz\right]  +
 \frac{\mathcal C_-}{(p_z)^{1/2}}\exp\left[  -\frac{i}{\hbar}\int^zp_zdz\right] .
\label{WKBin}
\end{eqnarray}
On the right hand of $z_+$, there is only an outgoing wave
\begin{eqnarray}
\frac{\mathcal T}{(p_z)^{1/2}}\exp\left[
 \frac{i}{\hbar}\int^z_{z_+}p_zdz\right] ,
\label{WKBout}
\end{eqnarray}
The connection equations
can be solved by
\begin{equation}
\bar {\cal C}=0,~~
{\mathcal C_\pm}=e^{\pm i\pi/4}{\mathcal C}/2,~~ {\cal T}=
{\mathcal C}_+ \exp \left[ -\frac{1}{\hbar}\int_{z_-}^{z_+}\kappa_z dz\right].
\label{outa}
\end{equation}
The incident
flux density is
\begin{eqnarray}
j_z\equiv \frac{\hbar}{2m_ei}\left[\phi^*\partial_z\phi - (\partial_z\phi^*)\phi\right]
=\frac{p_z}{m_e}\phi^*\phi=\frac{|{\mathcal C}_+|^2 }{m_e},
\label{influx}
\end{eqnarray}
which can be written as
\begin{eqnarray}
j_z(z)=v_z(z)n_-(z),
\end{eqnarray}
where
$v_z(z)=p_z(z)/m_e$
is the velocity
and $n_-(z)=\phi^*(z)\phi(z)$
the density of the incoming particles.
Note that the $z$-dependence
of
$v_z(z)$ and
$n_-(z)$
cancel each other.
By analogy,
the outgoing flux  density is
$|{\mathcal T}|^2 /m_e$.

\subsection{Rate of pair production}
From the above
considerations we obtain
for the transmission probability
\begin{eqnarray}
W_{\rm WKB}\equiv \frac{\rm transmitted ~~ flux}{\rm incident ~~ flux }
\label{wdefine}
\end{eqnarray}
the simple exponential
\begin{eqnarray}
W_{\rm WKB}(p_\perp,{\mathcal E}) &= & \exp\left[-
{\frac{2}{\hbar}}\int_{z_-}^{z_+} \kappa _zdz\right].
\label{tprobability1}
\end{eqnarray}
In order to derive
 from
(\ref{wdefine})
the total rate of pair production in the electric field we must multiply it
with the incident particle flux density
at the entrance $z_-$ of the tunnel.
The particle velocity
at that point is
 $v_z=\partial \E /\partial p_z$, where
the relation between $\E $ and $z_-$ is given by
 Eq.~(\ref{crosspoint+-}):
\begin{equation}
-1= \frac{{\mathcal E}-V(z_-)}{{\sqrt{(c{p}_\perp) ^2+m_e^2c^4}}}.
\label{@Entr}\end{equation}
This must be multiplied
with the particle density
which is given by the phase space density
$d^3p/(2\pi \hbar )^3$.
The incident flux density at the tunnel entrance is therefore
\begin{eqnarray}
j_z(z_-)=
\Ds\int  \frac{\partial \E}{\partial p_z}
\frac{d^2{p}_\perp}{(2\pi\hbar)^2}
\frac{d{p}_z}{2\pi\hbar} =
\Ds\int  \frac{ d\E}{{2\pi\hbar}}
\frac{d^2{p}_\perp}{(2\pi\hbar)^2},
\label{nflux}
\end{eqnarray}
and the extra factor $\Ds$
is equal to $2$ for electrons with two spin orientations \cite{remark}.

It is useful to change
 the variable of integration from  $z$ to $\yy(z)$
defined by
\begin{equation}
\yy({p}_\perp,\E;z)\equiv \frac{{\mathcal E}-V(z)}{{\sqrt{(c{p}_\perp) ^2+m_e^2c^4}}},
\label{y(x)}
\end{equation}
and to introduce the
notation for
the electric field
$E({p}_\perp,\E;\yy)\equiv E[\bar z({p}_\perp,\E;\yy)]$, where $\bar z(
{p}_\perp,\E;\yy)$ is the inverse function
of
(\ref{y(x)}),
the equations
in (\ref{crosspoint+-})
reduce to
\begin{equation}
\yy_-({p}_\perp,\E;z_-) =-1,~~~~
\yy_+({p}_\perp,\E;z_+) = +1.
\label{y+-}
\end{equation}
In terms of the variable $\zeta$,
the WKB transmission probability (\ref{tprobability1})
can be rewritten as
\begin{eqnarray}
W_{\rm WKB}(p_\perp,{\mathcal E}) &= & \exp\left\{ -\frac{2m^2_ec^3}{e \hbar E_0}
  \left[1+\frac{(c{p}_\perp) ^2}{m_e^2c^4}\right]
\int^{1}_{-1} d\yy\frac{\sqrt{1-\yy^2}}{ E({p}_\perp,\E;\yy)/E_0}\right\}.
\label{wwkbp}
\end{eqnarray}
Here we have introduced a standard field
strength $E_0$
to make the integral in the exponent
dimensionless, which we abbreviate
 by
\begin{eqnarray}
G(p_\perp,{\mathcal E})\equiv \frac{2}{\pi}\int^{1}_{-1} d\yy\frac{\sqrt{1-\yy^2}}{E({p}_\perp,\E;\yy)/E_0}.
\label{gf}
 \end{eqnarray}
The first term in the exponent
of (\ref{wwkbp})
is equal to $2E_c/E_0$, where
\begin{equation}
E_c\equiv m_e^2c^3/e\hbar
\label{@CRIT}\end{equation}
is the {\em critical field strength\/} which creates a pair over
two Compton wavelengths
$2 \lambda _C=2\hbar /m_ec$.

At the semiclassical level,
tunneling takes place only
if the potential
height is larger than $2m_ec^2$
and for energies $\E$
for which there are two real turning points $z_\pm$.
The total tunneling rate is obtained by integrating
over all
incoming momenta and the
total area
 $V_\perp=\int dxdy$
of the incoming flux. The WKB-rate per area is
\begin{eqnarray}
\frac{\Gamma _{\rm WKB}}{V_\perp}&=&
\Ds \int\frac{ d\E}{2\pi \hbar }
\int\frac{d^2{p}_\perp}{(2\pi\hbar)^2}
W_{\rm WKB}(p_\perp,{\mathcal E}).
\label{gxy0}
\end{eqnarray}
 Using the
relation following from (\ref{@Entr})
\begin{equation}
d{\mathcal E}=eE(z_-)dz_-,
 \label{@REL}\end{equation}
we obtain the alternative expression
\begin{eqnarray}
\frac{\Gamma _{\rm WKB}}{V_\perp}&=&
\Ds \int \frac{dz_-}{2\pi \hbar }
\int\frac{d^2{p}_\perp}{(2\pi\hbar)^2}
eE(z_-)
W_{\rm WKB}(p_\perp,{\mathcal E}(z_-)),
\label{gxy}
\end{eqnarray}
 where
${\mathcal E}(z_-)$ is obtained by solving the differential equation
(\ref{@REL}).

The integral over $p_\perp$
cannot be done
exactly.  At the semiclassical level,
this is fortunately not necessary.
Since $E_c$ is proportional
to $1/\hbar $, the exponential
in (\ref{wwkbp})
restricts the transverse momentum ${p}_\perp$
to be small of the order
of
$ \sqrt{\hbar }$,
so that the integral
in (\ref{gxy})
may be calculated from
an expansion
of $G(p_\perp,{\mathcal E})$ up to the order ${p}_\perp^2$:
\begin{eqnarray}
G(p_\perp,{\mathcal E})
\simeq \frac{2}{\pi}\int^{1}_{-1} d\yy\frac{\sqrt{1-\yy^2}}{ E(0,\E;\yy)/E_0}
\left[1-\frac{1}{2}\frac{d E({ 0},\E,\yy)/d\yy}{ E({ 0},\E,\yy)}\yy \,\delta +\dots
\right]=
G(0,{\mathcal E})+
G_ \delta (0,{\mathcal E}) \delta+\dots ,
\label{gfhbar}
\end{eqnarray}
where
$\delta\equiv  \delta (p_\perp)\equiv (c{p}_\perp) ^2/(m_e^2c^4)$,
and
\begin{eqnarray}
G_ \delta (0,{\mathcal E}) &\equiv &-\frac{1}{\pi}\int^{1}_{-1}
d\yy\frac{\yy\sqrt{1-\yy^2}}{ E^2(0,\E;\yy)/E_0} E'(0,\E;\yy)\nonumber\\
&=&-\frac{1}{2}
G(0,{\mathcal E})+\frac{1}{\pi}\int^{1}_{-1} d\yy\frac{\yy^2}{\sqrt{1-\yy^2}}
\frac{d\yy}{ E(0,\E,\yy)/E_0}.
\label{ghbar}
\end{eqnarray}
We can now perform the integral over ${\bf p}_\perp$
in (\ref{gxy})
approximately
as follows:
\begin{eqnarray}
\int \frac{d^2p_\perp}{(2\pi \hbar )^2}
e^{-\pi({E_c}/E_0)(1+ \delta )[
G (0,{\mathcal E})
+G_ \delta (0,{\mathcal E}) \delta ]}
&\approx&
\frac{m_e^2c^2}{4\pi\hbar ^2}\int_0^\infty d \delta  \,
e^{-\pi({E_c}/E_0)[
G (0,{\mathcal E}) + \delta
\tilde G (0,{\mathcal E})
}
\nonumber \\&=&
\frac{eE_0}{4\pi^2\hbar c\tilde
G (0,{\mathcal E})
}
e^{-\pi({E_c}/E_0)
G (0,{\mathcal E})},
\label{tranp}
\end{eqnarray}
where
\begin{equation}
\tilde
G (0,{\mathcal E})
\equiv G (0,{\mathcal E})
+G_ \delta  (0,{\mathcal E})
.
\end{equation}
%

The electric fields $E(p_\perp,\E;\yy)$
 at the tunnel entrance
$z_-$
in the prefactor of (\ref{gxy}) can
be expanded
similarly to first order in $ \delta $.
If $z_-^0$ denotes the solutions of
(\ref{@Entr})
at $p_\perp=0$,
we see that
for small $\delta$:
\begin{equation}
 \Delta z_-\equiv z_--z_-^0
\approx
\frac{m_ec^2}{E(z^0_-)}
\frac{\delta}{2} .
\end{equation}
so that
\begin{eqnarray}
E(z_-)\simeq
E(z^0_-)-
{m_ec^2}\frac{E'(z^0_-)}{E(z^0_-)}\frac{\delta}2.
\label{ehbar}
\end{eqnarray}
Here the extra term proportional to $\delta$
can be neglected in the semiclassical limit since it gives a contribution to the
prefactor of the order $\hbar $.
Thus we obtain
the WKB-rate (\ref{gxy}) of pair production
per unit area
\begin{eqnarray}
\frac{\Gamma _{\rm WKB}}{V_\perp}
\equiv \int dz \frac{\partial _z\Gamma _{\rm WKB}(z)}{V_\perp}
\simeq \Ds\int dz\frac{e^2 E_0 E(z)}{8\pi^3\hbar ^2c
\,\tilde G(0,{\mathcal E}(z))
}
e^{-\pi ( E_c/{E_0})G(0,{\mathcal E}(z))},
\label{pgxy3}
\end{eqnarray}
where
$z$ is short for  $z^0_-$.
At this point it is useful to
return from the integral
$\int dz_-
eE(z_-)
$  introduced
in (\ref{gxy}) to the
original energy integral
$\int d\E$
in
(\ref{gxy0}), so that the final result is
\begin{eqnarray}
\frac{\Gamma _{\rm WKB}}{V_\perp}\equiv
\int d\E \frac{\partial _{\E}\Gamma _{\rm WKB}(z)}{V_\perp}
\simeq \Ds
\frac{
eE_0
}{4\pi^2\hbar c}
\int \frac{d\E}{2\pi \hbar }
\frac{1}{
\tilde G(0,{\mathcal E})
}
e^{-{\pi ( E_c/E_0)G(0,{\mathcal E})}},
\label{pgwk1}
\end{eqnarray}
where ${\mathcal E}$-integration is over all crossing energy-levels.

These formula can be approximately applied to the 3-dimensional case of electric fields ${\bf E}(x,y,z)$ and potentials $V(x,y,z)$
at the points $(x,y,z)$ where
the tunneling length $a\equiv z_+-z_-$ is much smaller than the variation lengths $\delta x_\perp$
of electric potentials $V(x,y,z)$ in the $xy$-plane,
\begin{eqnarray}
\frac{1}{a} \gg \frac{1}{V}\frac{\delta V}{\delta x_\perp}.
\label{3dcon}
\end{eqnarray}
At these points $(x,y,z)$, we can arrange the tunneling path $dz$ and momentum $p_z(x,y,z)$
in the direction of electric field,
corresponding perpendicular area $d^2V_\perp\equiv dxdy$ for incident flux and perpendicular
momentum ${\bf p}_\perp$. It is then approximately reduced to a one-dimensional problem in the region of size
$~{\cal O}(a)$ around these points.
The
surfaces
$z_-(x_-,y_-,\E)$
and
$z_+=(x_+,y_+,\E)$
 assciated with the
classical turning points
are determined by
Eqs.~(\ref{y(x)}) and
Eqs.~(\ref{y+-}) for a given energy $\E$.
The WKB-rate of pair production
(\ref{pgxy3}) can then be expressed
as an volume integral
over the
rate density per volume element
\begin{eqnarray}
\Gamma _{\rm WKB}=
\int dxdydz
\frac{d^3\Gamma _{\rm WKB}}{dx\,dy\,dz}=
\int dtdxdydz
\frac{d^4N_{\rm WKB}}{dt\,dx\,dy\,dz}.
\label{3drate0}
\end{eqnarray}
On the right-hand side we have
found it useful to
rewrite the rate $
\Gamma _{\rm WKB}$ as the time derivative
of the number of pair creation events  $dN_{\rm WKB}/dt$, so that we obtain
an event density in four-space

\begin{eqnarray}
\frac{d^4N_{\rm WKB}}{dt\,dx\,dy\,dz}
\approx \Ds \frac{e^2 E_0 E(z)}{8\pi^3\hbar
\,\tilde G(0,{\mathcal E}(z))
}
e^{-\pi ( E_c/{E_0})G(0,{\mathcal E}(z))},
\label{3drate}
\end{eqnarray}
Here $x,y$ and $z$ are
related by the function $z=z_- (x,y,\E)$ which is
 obtained by solving (\ref{@REL}).

It is now useful
to observe
that the left-hand side of (\ref{3drate})
is a Lorentz-invariant quantity. In addition, it is
symmetric
under the exchange of time and $z$, and this symmetry will be exploited
in the next section to  relate pair production processes in
a $z$-dependent electric field $E(z)$
to those in a time-dependent field $E(t)$.

\comment{
The limits of integration are $\E\in (-1+ \sigma ,1- \sigma )$.}

\comment{
The result can be extended
to the production of $n$ particle-antiparticle pairs, for which the
 WKB-probability is given by
\begin{eqnarray}
\sum_{n=1}^{\infty}-\frac{(-1)^n}{n}\left[W_{\rm WKB}(p_\perp,{\mathcal E})\right]^n,
\label{nwkb}
\end{eqnarray}
and the total WKB-rate per volume
\begin{eqnarray}
\frac{\Gamma _{\rm WKB}[{\mathcal E}(z_-),z_-]}{V}&=&\Ds
\int\frac{d^2{p}_\perp}{(2\pi\hbar)^3}
eE(z_-)
\sum_{n=1}^{\infty}\frac{1}{n}\left[W_{\rm WKB}(p_\perp,{\mathcal E})\right]^n .
\label{ngwkb}
\end{eqnarray}
}

        Attempts to go beyond the WKB results
(\ref{pgxy3}) or
(\ref{pgwk1}) require a great amount of work.
Corrections will come from three sources:
\begin{enumerate}
\item[I] from the
higher terms
of order in~$ (\hbar)^n$ with $n>1$
 in the the expansion (\ref{4.13})
solving
the Riccati equation
(\ref{n4.5}).
\item[II]
from  the perturbative evaluation of the
integral over ${\bf p}_\perp$ in Eqs. (\ref{gxy0})
 or
(\ref{gxy})
when going beyond the Gaussian approximation.
\item[III]
from perturbative
 corrections
to the Gaussian
energy integral (\ref{pgwk1})
or the corresponding $z$-integral (\ref{pgxy3}).
\end{enumerate}
All these corrections contribute terms of higher order in $ \hbar$.

\label{@RATE}

\subsection{Including a Smoothly Varying ${\bf B}(z)$-Field  Parallel to ${\bf E}(z)$}

The above results can easily be extended to allow for the
presence of a constant
magnetic field
 ${\bf B}$ parallel to ${\bf E}(z)$.
Then
the wave
function factorizes into a Landau state and
a
spinor function first calculated by
Sauter \cite{sauter}.
In the WKB approximation,
the energy spectrum
 is still given by Eq.~(\ref{energyl+-}),
but the squared transverse momenta $p_\perp^2$ is
quantized and must be
 replaced
by discrete values corresponding to
the Landau energy levels. From the known nonrelativistic levels
for the Hamiltonian $p_\perp^2/2m_e$ we extract
 immediately the replacements
\begin{equation}
c^2p_\perp^2 =
2m_ec^2 \times\left( \frac{ p_\perp^2}{2m_e}\right)
\longrightarrow
2m_ec^2 \times\left[ {\hbar \omega _L}
\left(n\!+\!\frac{1}{2}\!+\!{g}\sigma\right)
\right]
,~n=0,1,2,\cdot\cdot\cdot~ ,
\label{landaulevel}\end{equation}
where
$g= 2+ \alpha /\pi+\dots $ is the anomalous magnetic moment of the electron,
$\omega _L=eB/m_e c$ the Landau frequency,
with $\sigma=\pm 1/2$ for spin-$1/2$ and $\sigma=0$ for
 spin-$0$, which are eigenvalues of the Pauli matrix
$\sigma_z$. The quantum number $n$
characterizing the
Landau levels
counts the
levels of the harmonic oscillations in the plane orthogonal to
the $z$-direction.
Apart from the replacement
(\ref{landaulevel}), the WKB calculations
remain the same.
Thus
we must only
replace the
integration over the transverse momenta
$\int d^2{ p}_\perp/(2\pi\hbar)^2$
in Eq.~(\ref{tranp}) by the sum
over all Landau levels with the degeneracy
$ eB/(2\pi\hbar c)$. Thus, the right-hand side becomes
\begin{equation}
  \frac{eB}{2\pi\hbar c}e^{-\pi({E_c}/E_0)
G (0,{\mathcal E})}
\sum_{n,\sigma}
e^{-\pi (B/E_0)(n+1/2+g\sigma)
\tilde G (0,{\mathcal E})
}.
\label{tprobability2h}  \!\!\!\!\!\!\!
\end{equation}
The result is,
for spin-0 and spin-1/2:
\begin{eqnarray}
\frac{eE_0}{4\pi^2\hbar c\tilde G (0,{\mathcal E})}
e^{-\pi({E_c}/E_0)G (0,{\mathcal E})}
f_{0,1/2}(
 B
\tilde G (0,{\mathcal E})/E_0)
\label{wkbehfermion2}\end{eqnarray}
where
\begin{eqnarray}
f_{0}(x)\equiv
\frac{\pi x}{\sinh \pi x},\ \ \  \
f_{1/2}(x)\equiv
2\frac{ \pi x }{\sinh \pi x}
{ \cosh \frac{\pi gx}2}
\label{wkbehboson}\end{eqnarray}
In the limit $B\rightarrow 0$,
Eq.~(\ref{wkbehboson})
reduces
 to Eq.~(\ref{tranp}).

The result remains approximately valid if the magnetic
field has a smooth $z$-dependence varying little
over a Compton wavelength $ \lambda _e$.

In the following we shall focus only on nonuniform electric fields
without a magnetic field.

\section{Time-dependent electric fields}\label{time}

The above semiclassical considerations
can be applied with little change
to the different physical situation
in which the
electric field along the $z$-direction depends only on time
rather than  $z$. Instead of the time $t$ itself we shall prefer working with
the
zeroth length coordinate $x_0=ct$, as usual in relativistic calculations.
As an intermediate step consider for a moment a
vector potential
\begin{equation}
A_\mu=(A_0(z),0,0,A_z(x_0)),
\end{equation}
with the
electric field
\begin{equation}
E=-\partial _zA_0(z)-\partial_0A_z(x_0),~~~~x_0\equiv ct.
\end{equation}
The associated Klein-Gordon equation  (\ref{@KG0})
reads
\begin{eqnarray}
\left\{ \left[ i\hbar \partial _0+\frac{e}{c}A_0(z)\right] ^2
+\hbar ^2\partial _{{\sbf x}_\perp}^2
-\left[ i\hbar \partial _z+\frac{e}{c}A_z(x_0)\right] ^2-m_e^2c^2\right\}
\phi(x)=0.
\label{@KG}\end{eqnarray}
The previous discussion
was valid under the assumption
$A_z(x_0)=0$, in which case the ansatz
$\phi(x)=e^{-i{\cal E}t/\hbar }e^{i{\sbf p}_\perp {\sbf x}_\perp}
\phi_{{\sbf p}_\perp,\E}(z)$ led
to the
field equation (\ref{KG}).
For the present discussion
it is useful to
write the ansatz as
$\phi(x)=e^{-ip_0x_0/\hbar }e^{i{\sbf p}_\perp {\sbf x}_\perp/\hbar } \phi_{{\sbf p}_\perp,p_0}(z)$
with
$p_0=\E/c$, and Eq.~(\ref{KG})
in the form
\begin{eqnarray}
\left\{\frac{1}{c^2}\left[\E-e\int^z dz'\,E(z')\right]^2-p_\perp^2
-m_e^2c^2
+\hbar ^2\frac{d^2}{dz^2}
\right\}  \phi_{{\sbf p}_\perp,p_0}(z)=0.
\label{KG2}\end{eqnarray}

Now we assume the electric field
to depend only on $x_0=ct$. Then
the
ansatz
$\phi(x)=e^{ip_zz/\hbar }e^{i{\sbf p}_\perp {\sbf x}_\perp/\hbar } \phi_{{\sbf p}_\perp,p_z}(x_0)$
leads to the field equation
\begin{eqnarray}
\left\{-\hbar ^2\frac{d^2}{d{x_0}^2}-p_\perp^2
-m_e^2c^2-\left[-p_z-\frac{e}{c}\int^{x_0} dx'_0 E(x'_0) \right]^2\right\}  \phi_{{\sbf p}_\perp,p_z}(x_0)=0.
\label{KG3}\end{eqnarray}

\comment{
The transition
from $z$- to
the time-dependent
electric fields
is done by performing
simultaneously  a Wick rotation
from $x_0$ to the Euclidean variable $x_0^E\equiv x_4\equiv ict=ix_0$, and an anti-Wick
rotation  of the Euclidean coordinate
$z$ to the time-like
coordinate $z^M=-iz$.
The superscript $M$ stands for Minkowski
to indicate that the space $x,y,z^M,x_4$
has now the Minkowski metric.
The transformed
field equation reads
\begin{eqnarray}
\left\{\hbar ^2\frac{d^2}{d{x_4}^2}-p_\perp^2
-m_e^2c^2-\left[p_z-eA_z(x_0)/c \right]^2\right\}  \phi_{{\sbf p}_\perp,p_0}(x_0)=0.
\label{KG4}\end{eqnarray}
amounts to assuming
the presence of only
a nonzero second part with $A_z=A_z(x_0)$,
rather than the
first part in Eq.~(\ref{@VEQ}).
Now the
 other components
$A_t,\,A_x,\,A_x$
are assumed to vanish.
To keep the analogy with (\ref{@VEQ})
as close as possible, we shall denote $-eA_z(x_0)$ by $V(x_0)/c$
so that the canonical momentum in the $z$-direction
can be written as
\begin{equation}
P_z(x_0)=p_z(x_0)-V(x_0)/c.
\end{equation}
The pair production
is now a consequence
of a force pulse
$\delta p_z(x_0)\simeq -eE(x_0)\delta x_0$
pulling a negative-energy electron
across the gap $2 m_ec^2$.
}

If we compare Eq.~(\ref{KG3})
with (\ref{KG2})
we realize that
one arises from the other by interchanging
\begin{equation}
z\leftrightarrow x_0,~~~p_\perp\rightarrow ip_\perp, ~~~c\rightarrow ic,~~~E\rightarrow -iE.
\label{trantable}\end{equation}
With these exchanges
we may easily calculate the
decay rate of the vacuum
caused by a time-dependent electric field $E(x_0)$
using the above-derived formulas.

\section{Applications}

We now
apply Formulas~(\ref{pgwk1})
or (\ref{pgxy3})
to various external field configurations
capable of producing electron-positron pairs.

\subsection{Step-like electric field}

First we check our result
 for the original
case of a constant electric field
$E(z)\equiv eE_0$
where the
potential energy is the linear function  $V(z)=-eE_0z$.
Here the function (\ref{gf}) becomes trivial
\begin{eqnarray}
G(0,\E)=\frac{2}{\pi}\int^{1}_{-1} d\yy\sqrt{1-\yy^2}=1,\quad G_\delta(0,\E)=0,
\label{gfc}
\end{eqnarray}
which is independent of $\E$ (or $z_-$).
The WKB-rate
 for pair-production
per unit time and volume is found from Eq.~(\ref{pgxy3})
to be
\begin{equation}
 \frac{\Gamma^{\rm EH} _{\rm WKB}}{V}\simeq \Ds\frac{e^2 E_0^2}{ 8\pi^3\hbar^2 c}
e^{-\lfrac{\pi E_c}{E_0}}.
\label{xy2}
\end{equation}
where $V\equiv dz_- V_\perp $. This expression contains
the exponential
$e^{-\pi E_c/E_0}$ found by Sauter \cite{sauter}, and
the correct prefactor
as
calculated by
Heisenberg and Euler \cite{euler},
and by Schwinger \cite{schwinger}.

In order to apply the translation table (\ref{trantable}) to obtain the analogous result for the constant electric field in time,
we rewrite Eq.~(\ref{xy2}) as
\begin{equation}
 \frac{dN_{\rm WKB}}{dx_0V}\simeq \Ds\frac{e^2 E_0^2}{ 8\pi^3\hbar^2 c^2}
e^{-\lfrac{\pi E_c}{E_0}},
\label{txy2}
\end{equation}
where $dN_{\rm WKB}/dx_0=\Gamma^{\rm EH} _{\rm WKB}/c$ and $N_{\rm WKB}$ is
the number of pairs produced. Applying the translation table (\ref{trantable}) to Eq.~(\ref{txy2}),
one obtains the same formula as Eq.~(\ref{xy2}).

\subsection{Sauter electric field}

Let us now consider the
nontrivial Sauter electric field
localized within finite slab in the $xy$-plane with the width $\ell $ in the $z$-direction.
A field of this type
can be produced, e.g.,
between
two opposite charged conducting plates.
The electric field $E(z)\hat{\bf z}$
in the $z$-direction
and the associated
potential energy $V(z)$ are given by
\begin{eqnarray}
E(z)=E_0/{\rm cosh}^2\left({z}/{\ell }\right),~~~~~\label{sfield}
V(z)&=&- \sigma\, m_ec^2\tanh\left({z}/{\ell }\right),
\label{sauterv}
\end{eqnarray}
where
\begin{equation}
\sigma\equiv
eE_0
\ell /m_ec^2=(\ell /\lambda_C)(E_0/E_c).
\label{@gamm}\end{equation}
\comment{
and $\lambda_C=\hbar /m_ec$ is the electron Compton length.}
From now on we shall use natural units
in which energies are
measured
in units of $m_ec^2$.
Figure~\ref{sauterf}
shows
the positive and negative-energy spectra
${\mathcal E}_\pm(z)$ of Eq.~(\ref{energyl+-}) for $p_z=p_\perp=0$ to show the energy-gap and
energy-level crossings.
From Eq.~(\ref{crosspoint+-}) we find the classical
turning points
\begin{equation}
z_\pm=   \ell   ~
{\rm arc tanh}\frac{\E\pm\sqrt{1+ \delta }}{ \sigma }=
\frac{\ell }2\ln\frac{ \sigma+  {\mathcal E}\pm \sqrt{1+ \delta }}{
 \sigma-  {\mathcal E} \mp
 \sqrt{1+ \delta }}.
\label{scrossing}
\end{equation}
Tunneling is possible
for all
energies satisfying
\begin{equation}
- \sqrt{1+ \delta }+ \sigma \ge  {\mathcal E} \ge  \sqrt{1+ \delta }-  \sigma ,
\label{crossingregime}
\end{equation}
for the strength parameter
$ \sigma >\sqrt{1+ \delta }$.

\comment{
The electric field at the tunnel entrance $z_-$ is given by
\begin{equation}
E(z_-)=E_0[1-\tanh^2(z_-/2\ell )]  .
\label{sey0}
\end{equation}
}
We may invert
Eq.~(\ref{y(x)}) to find the relation between $\zeta$ and $z$:
\begin{equation}
z=z(p_\perp,\E;\zeta)= \ell   \,
{\rm arc tanh}\frac{\E+ \zeta \sqrt{1+ \delta }}
{ \sigma }=\frac{\ell }2\ln\frac{ \sigma+  {\mathcal E}+\zeta \sqrt{1+ \delta }
}{
 \sigma-  {\mathcal E} -
  \zeta \sqrt{1+ \delta }}.
\label{@zEqu}\end{equation}
In terms of  the function
$z(p_\perp,\E;\zeta)$,
the equation
(\ref{scrossing}).
reads simply
$z_\pm=z(p_\perp,\E;\pm1)$.

Inserting (\ref{@zEqu})
into the equation
for $E(z)$ in Eq.~(\ref{sauterv}),
we obtain
\begin{equation}
E(z)= E_0\left[1-\left(\frac{ \yy\sqrt{1+\delta}- {\mathcal E}}{ \sigma}\right)^2\right]
\equiv E(p_\perp,\E;\yy).
\label{sey}
\end{equation}
We now calculate
$G(0,{\mathcal E})$
 and $G_ \delta (0,{\mathcal E})$
of Eqs.~(\ref{gf}), (\ref{gfhbar}) and (\ref{ghbar}):
\begin{eqnarray}
G(0,{\mathcal E})&=&2 \sigma^2 - \sigma
\left[( \sigma -  {\mathcal E})^2 -1\right]^{1/2}- \sigma
\left[( \sigma +  {\mathcal E})^2 -1\right]^{1/2},
\label{sgf}
\end{eqnarray}
and
\begin{eqnarray}
G (0,{\mathcal E})+G_\delta (0,{\mathcal E})&=&\frac{\sigma}{2}\left\{
\left[( \sigma -  {\mathcal E})^2 -1\right]^{-1/2}+
\left[(\sigma + {\mathcal E})^2 -1\right]^{-1/2}\right\}.
\label{pgsgf}
\end{eqnarray}
Substituting the functions $G (0,{\mathcal E})$ and $G_\delta (0,{\mathcal E})$
into
Eqs.~(\ref{pgxy3}) and (\ref{pgwk1}),
we obtain the general expression
for the pair-production rate per volume slice at a given
tunnel entrance  point $z_-(\E)$ or the associated energy $\E(z_-)$.
The pair-production rate per area is obtained by integrating
over all slices permitted by the energy inequality (\ref{crossingregime}).

\begin{figure}[tbhp]

\unitlength1mm
\def\fsz{\footnotesize}
\def\ssz{\scriptsize}
\def\tsz{\tiny}
\def\dst{\displaystyle}
\def\pu#1#2{\put(#1,#2){\emmoveto}}
\def\pd#1#2{\put(#1,#2){\emlineto}}
\begin{picture}(105.64,28.645)
\def\dst{\displaystyle}
\def\fsz{\footnotesize}
\put(-10,0){\includegraphics[width=4.6cm]{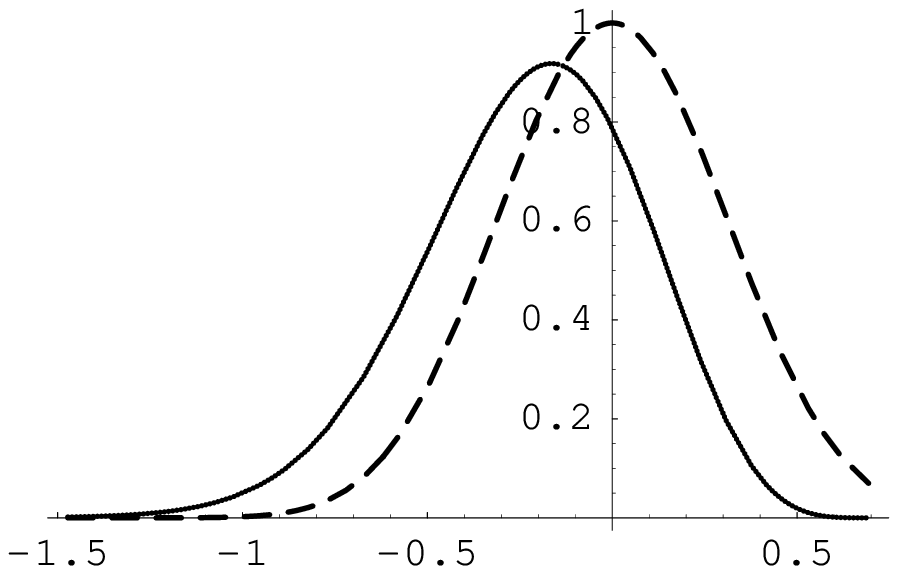}}
\put(70,.5){\includegraphics[width=3.8cm]{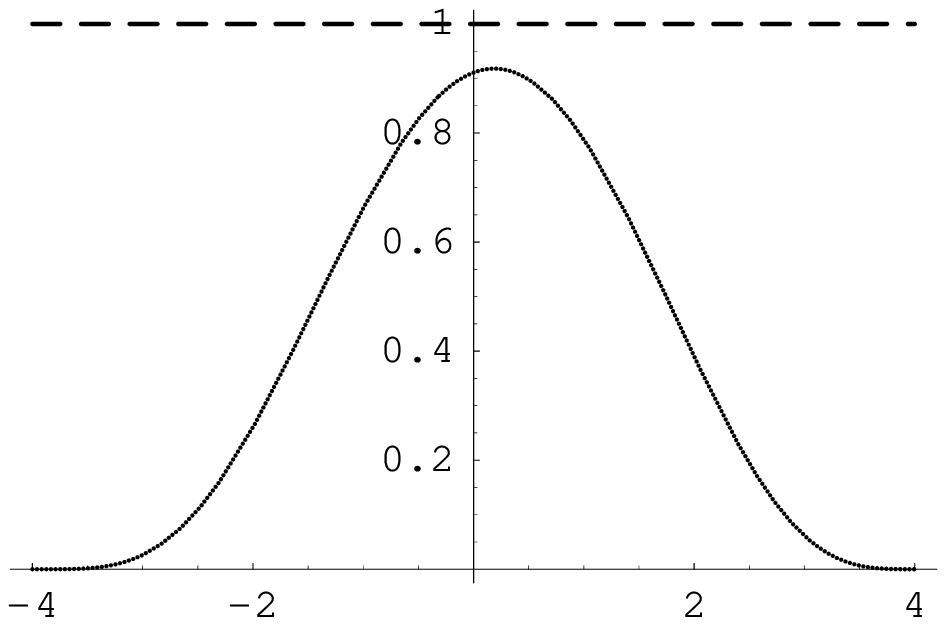}}
\put(107,2){\fsz $z/\ell$}
\put(32,2){\fsz ${\cal E}/mc^2$}
\put(24,21){\tsz $\Gamma ^{\rm EH}_ {\rm WKB}[ E_0\rightarrow E(z)]
/\Gamma^{\rm EH} _ {\rm WKB}$}
\put(-15,14){\tsz $\partial _z\Gamma _ {\rm WKB}/\Gamma^{\rm EH} _ {\rm WKB}$}
\put(97,14){\tsz $\partial _{\E}\Gamma _ {\rm WKB}/\Gamma^{\rm EH} _ {\rm WKB}$}
\end{picture}
\caption[]{We plot the slice dependence of the integrand in the tunneling rate (\ref{pgxy3})
for the Sauter potential  (\ref{sauterv}): left,
as a function of the tunnel entrance $z$ (compare
with numeric results plotted in Fig. 1 of Ref. \cite{gk2005}); right, as a function of
the associated energy $\cal E$, which is normalized by the Euler-Heisenberg rate (\ref{xy2}).
The dashed curve in left figure shows the Euler-Heisenberg  expression (\ref{xy2})
evaluated for the $z$-dependent field $E(z)$ to illustrate the nonlocality of the
production rate. The dashed curve in right figure shows the Euler-Heisenberg  expression (\ref{xy2})
is independent of energy-level crossing $\E$. The dimensionless parameters are $ \sigma =5,\,E_0/E_c=1 $. }
\label{TUNNf}
\end{figure}

In Fig. \ref{TUNNf}
we show the slice dependence of the integrand in the tunneling rate
(\ref{pgxy3}) for the Sauter potential (\ref{sauterv})
and compare it with the
constant-field expression
(\ref{xy2})
of
Euler and Heisenberg, if this is evaluated at the $z$-dependent
electric field $E(z)$.
This is done once as a function of the tunnel entrance point $z$
and once as a function of the associated energy $\cal E$.
On each plot, the difference between the two curves
illustrates the nonlocality of the tunneling process \cite{NONL}.

The integral is dominated by
the
 region around $\E\sim 0$,
where the tunneling length is shortest [see Fig.~\ref{sauterf}] and tunneling probability is largest. Both functions
$G (0,{\mathcal E})$
and $G_\delta (0,{\mathcal E})$
have a symmetric peak at $\E=0$.
Around the peak
they can be expanded in powers of $
{\mathcal E}$  as
\begin{eqnarray}
G (0,{\mathcal E})=2 [
\sigma^2 - \sigma(\sigma^2 -1)^{1/2}]+\frac{\sigma}{(\sigma^2 -1)^{3/2}}
{\E^2}
 + {\cal O}(\E^4)
=
G_0( \sigma )+ \frac{1}{2}G_2( \sigma )\,
{\E^2} +{\cal O}(\E^4)
,
\label{exppsgf}
\end{eqnarray}
and
\begin{eqnarray}
G (0,{\mathcal E})+
G_ \delta (0,{\mathcal E})
 =
\frac{\sigma}{ ( \sigma ^2 -1)^{1/2}}+ \frac{1}{2}
\frac{(1 +2 \sigma^2)}{(\sigma ^2 -1)^{5/2}}
{\E^2} +{\cal O}(\E^4)
=
\overline G_0( \sigma )+ \frac{1}{2}\overline G_2( \sigma )\,
{\E^2} +{\cal O}(\E^4) .
\label{exppgsgf}
\end{eqnarray}
The exponential $
e^{-\lfrac{\pi G(0,{\mathcal E}) E_c}{E_0}}$ has a Gaussian peak
around $\E=0$ whose width
is of the order of $1/E_c\propto \hbar$.
This implies that in the semiclassical limit,
we may perform only a Gaussian integral
 and
neglect the $\E$-dependence of the prefactor
in (\ref{pgwk1}).
Recalling that
$\E$ in this section
is in natural units with $m_ec^2=1$,
we must replace $\int d\E$ by $ m_ec^2 \int d\E$ and can perform
the integral over $\E$
   approximately
as follows
\begin{eqnarray}\!\!\!\! \!\!
\frac{\Gamma _{\rm WKB}}{V_\perp}
\!\simeq \!
 \Ds
\frac{
eE_0
m_ec^2 }{4\pi^2\hbar c}
\frac{1}{\overline G_0}
e^{-{\pi (E_c/E_0) G_0 }}
\!\!\int\! \frac{d\E}{2\pi \hbar }
e^{-{\pi (E_c/E_0)G''_0\, }\E^2/{2}}
\!=\! \Ds
\frac{
eE_0
}{4\pi^2\hbar c}
\frac{1}{\overline G_0} \frac{e^{-{\pi(E_c/E_0) G_0 }}}{
2\pi \hbar  \sqrt{G_0''E_c/2E_0} }
.
\label{pgwk4}
\end{eqnarray}
For convenience,
we have extended
the limits of integration
over $E$
from the interval $(-1+ \sigma ,1- \sigma )$ to
$(-\infty,\infty)$.
This introduces
exponentially small errors
and can be ignored.

Using
the relation (\ref{@gamm})
we may replace $
eE_0
m_ec^2/\hbar c$ by $ e^2 E_0^22\ell / \sigma $,
and obtain
\begin{eqnarray}
\frac{\Gamma _{\rm WKB}[\rm total]}{V_\perp \ell }
\simeq \Ds\frac{e^2 E^{2}_0 }{8\pi^3\hbar^2 c}
 \sqrt{\frac{E_0}{E_c}}\frac{(\sigma^2-1)^{5/4}}{\sigma^{5/2}}
e^{-\lfrac{\pi G_0( \sigma ) E_c}{E_0}}
.
\label{expgwkb1}
\end{eqnarray}
This approximate result agrees \cite{DS}
 with that
\comment{
was}
obtained before
with a different, somewhat more complicated technique
proposed by
Dunne and Schubert
\cite{ds2005} after the
fluctuation determinant
was calculated exactly
in \cite{dwgs2006}
 with the help of
the Gelfand-Yaglom method
following Ref.~\cite{PI4}.
The advantage of knowing the exact
fluctuation determinant
could not, however,
be fully exploited
since the remaining integral
was calculated only in the saddle point approximation.
The rate (\ref{expgwkb1}) agrees  with
\comment{
the
result (4.7) of
Dunne et al. \cite{dwgs2006}
and with}
the
leading term of the expansion
(42)
of Kim and Page
\cite{kp2007}.
Note that the higher expansion terms
calculated by the latter authors do not yet lead to
proper higher-order results since
they are only of type II and III in the list
after Eq. (\ref{3drate}).
The
 terms
of equal order in $\hbar$
in the expansion
 (\ref{4.13}) of the solution of the Riccatti equation
are still missing.

Using the translation table (\ref{trantable}), it is straightforward to obtain
the pair-production rate of the Sauter-type of  electric field depending on time
rather than space.
According to the translation table (\ref{trantable}), we have
to replace
${\ell}\rightarrow c{\mathcal T}$,
where ${\mathcal T}$ is the characteristic time over which
the electric field acts---the analog of $\ell$
in (\ref{sauterv}).
Thus the field (\ref{sauterv}) becomes
\begin{eqnarray}
E(t)=E_0/{\rm cosh}^2\left({t}/{\mathcal T}\right),~~~~~\label{tsfield}
V(t)&=&- \tilde\sigma\, m_ec^2\tanh\left({t}/{\mathcal T }\right).
\label{tsauterv}
\end{eqnarray}
According to the same table we must also replace $\sigma\rightarrow i\tilde\sigma$,
where
\begin{equation}
\tilde\sigma\equiv
eE_0
{\mathcal T} /m_ec.
\label{t@gamm}
\end{equation}
This brings
$G_0(\sigma)$ of Eq. (\ref{sgf})
to the form
\begin{eqnarray}
G_0(\sigma)\rightarrow G^t_0(\tilde\sigma)=
2 [\tilde\sigma(\tilde\sigma^2 - 1)^{1/2}-\tilde\sigma^2],
\label{gt0}
\end{eqnarray}
and yields the pair-production rate
\begin{eqnarray}
\frac{\Gamma^z _{\rm WKB}[\rm total]}{V_\perp {\mathcal T} }
\simeq \Ds\frac{e^2 E^{2}_0 }{8\pi^3\hbar^2 c}
 \sqrt{\frac{E_0}{E_c}}\left(\frac{\tilde\sigma^2+1}{\tilde\sigma^2}\right)^{5/4}
e^{-\lfrac{\pi G^t_0( \tilde\sigma ) E_c}{E_0}},
\label{texpgwkb1}
\end{eqnarray}
where $\Gamma^z _{\rm WKB}[\rm total]=\partial N_{\rm WKB}/\partial z$ is the number of pairs produced
per unit thickness in a spatial shell parallel to the $xy$-plane.
This agrees with Ref. \cite{dwgs2006}.

Note also that the constant-field result
(\ref{xy2})        of Euler and Heisenberg
cannot be deduced from (\ref{expgwkb1})
by simply taking the limit $\ell \rightarrow \infty$
as one might have
 expected.
The reason is
that
the saddle point approximation
(\ref{pgwk4}) to the integral
(\ref{pgwk1})
becomes invalid in this limit.
Indeed, if
 $ \ell \propto \sigma$ is large
in Eqs.~(\ref{sgf})
and (\ref{pgsgf}), these become
\begin{eqnarray}
G(0,{\mathcal E})\rightarrow
G (0,{\mathcal E})+G_\delta (0,{\mathcal E})
\rightarrow  \frac{1}{1-\E^2/ \sigma ^2},
\end{eqnarray}
and the integral in (\ref{pgwk1}) becomes approximately
\begin{eqnarray}
e^{-{\pi ( E_c/E_0)}}
\int _{- \sigma }^ \sigma \frac{d\E}{2\pi \hbar }
\left({1-\E^2/ \sigma ^2}\right)
e^{-{\pi ( E_c/E_0)(\E^2/ \sigma ^2)}}
\label{@XXX}\end{eqnarray}
For not too large $ \ell \propto \sigma$, the integral can be evaluated
in the leading Gaussian approximation
\begin{eqnarray}
\int _{-\infty }^ \infty \frac{d\E}{2\pi \hbar }
e^{-{\pi ( E_c/E_0)(\E^2/ \sigma ^2)}}=\frac{1}{2\pi \hbar } \sqrt{\frac{E_0}{E_c}}\sigma,
\end{eqnarray}
corresponding
to the previous result (\ref{expgwkb1})
for large-$ \sigma $.
For a constant field, however, where the integrands becomes flat,
the Gaussian approximation is no longer applicable. Instead we must
first
set
 $ \sigma \rightarrow \infty $ in
the integrand of (\ref{@XXX}), making it constant.
Then the integral
(\ref{@XXX})
 becomes
\cite{REMA2}
\begin{eqnarray}
e^{-{\pi ( E_c/E_0)}}2\sigma/2\pi \hbar =
 e^{-{\pi ( E_c/E_0)}}2\ell
eE_0
/m_ec^2\,2\pi \hbar .
\end{eqnarray}
Inserting this
into (\ref{pgwk1})
we recover
the constant-field result (\ref{xy2}). We must replace
$2\ell  $ by $L$ to comply with the
relation
(\ref{@REL})
from which we obtain
\begin{eqnarray}
\int d\E=\int dz
eE(z)
=
eE_0
\int dz/\cosh^2(z/\ell )=2\ell
eE_0
=L
eE_0
.
\nonumber
\end{eqnarray}

\begin{figure}[th]
\begin{center}
\begin{picture}(35.64,114.645)
\def\fsz{\footnotesize}
\def\ssz{\scriptsize}
\def\tsz{\tiny}
\def\dst{\displaystyle}\unitlength1mm
\put(-60,0){\includegraphics[width=6cm,clip]{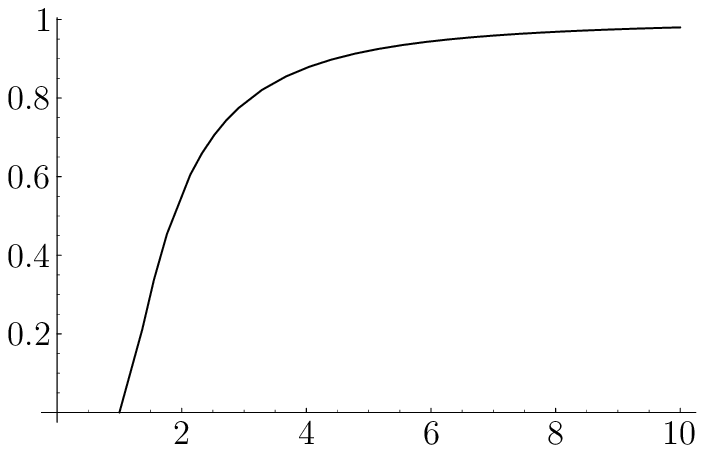}}
\put(10,-1){\includegraphics[width=6cm,clip]{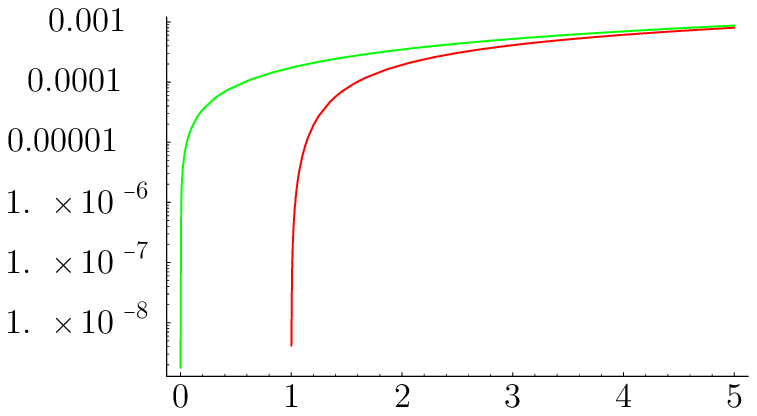}}
\put(-15,27.9){\footnotesize$R_{\rm rate}$}
\put(35,30.9){\footnotesize${\Gamma _{\rm WKB}[\rm total]}/{V_\perp \ell }$}
\put(-3,2.4){\footnotesize$ \sigma $}
\put(68,2.4){\footnotesize$ \sigma $}
\end{picture}
\end{center}
\caption{Left: Ratio $R_{\rm rate}$
defined in  Eq.~(\ref{compcs}) is plotted as function of
$\sigma$ in the left figure.
Right:
Number of pairs created in slab
of Compton width per area and time as functions of $\sigma$.
Upper curve is
for the constant field (\ref{xy2}), lower
for the Sauter field (\ref{expgwkb1})).
Both plots
are for $E_0=E_c$ and $\sigma=\ell /\lambda_C$.
}
\label{rratef}%
\end{figure}

In order to see the boundary effect on the pair-production rate,
we close this section with a comparison
between pair-production rates in the constant field (\ref{xy2}) and
Sauter field (\ref{expgwkb1}) for the same field strength $E_0$ in the volume
$V_\perp \ell $. The ratio $R_{\rm rate}$ of
pair-production rates (\ref{expgwkb1}) and (\ref{xy2}) in the volume $V_\perp \ell $ is defined as
\begin{eqnarray}
R_{\rm rate}= \sqrt{\frac{E_0}{E_c}}e^{\lfrac{\pi E_c}{E_0}}\frac{(\sigma^2-1)^{5/4}}{\sigma^{5/2}}
e^{-\lfrac{\pi G_0( \sigma ) E_c}{E_0}}.
\label{compcs}
\end{eqnarray}
The
soft boundary of the
Sauter field (\ref{sauterv}) reduces its
pair-production rate
with respect to the pair-production rate (\ref{xy2}) computed in a constant field
of width $L=2\ell$.
The reduction is shown quantitatively
in Fig.~\ref{rratef},
where curves are plotted for
the rates
(\ref{xy2}) and (\ref{expgwkb1}), and
and for their ratio (\ref{compcs}) at
$E_0=E_c$ and $\sigma=\ell/\lambda_C$ [recall (\ref{@gamm})].
We see that the reduction is
significant
if the
width of the field slab
shrinks to
the size
of a
 Compton wavelength
$ \lambda _C$.

\subsection{Constant electric field for $z>0$}

\begin{figure}[th]
\begin{center}
\begin{picture}(105.64,184.645)
\def\fsz{\footnotesize}
\def\ssz{\scriptsize}
\def\tsz{\tiny}
\def\dst{\displaystyle}\unitlength1mm
\put(-20,0){\includegraphics[width=8cm,clip]{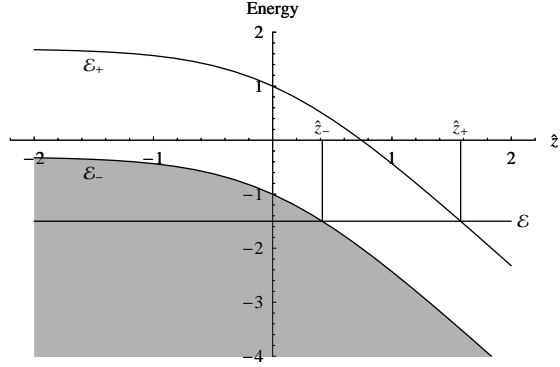}}
\end{picture}
\end{center}
\caption{Energies (\ref{energyl+-})
for a soft electric field step $E(z)$ of Eq.~(\ref{hfield}) and the potentials $V_\pm(z)$
(\ref{hv}) for $\sigma=5$.
Positive and negative-energies ${\mathcal E}_\pm(z)$
of Eq.~(\ref{energyl+-}) are plotted for $p_z= p_\perp =0$
as functions of $\hat z=z/\ell$. }%
\label{halfspace}%
\end{figure}

As a second application consider
an electric field
which is zero for $z<0$ and goes to $-E_0$
over a distance $\ell$
as follows:
\begin{eqnarray}
E(z)=-\frac{E_0}{2}\left[\tanh\left(\frac{z}{\ell}\right)+1\right],~~~\label{hfield}
V(z)= -\frac{\sigma}2 m_ec^2\left\{\ln\cosh\left(\frac{z}{\ell}\right)+\frac{z}{\ell}\right\},
\label{hv}
\end{eqnarray}
where $ \sigma \equiv eE_0\ell/m_ec^2$.
In Fig.~\ref{halfspace}, we have plotted the positive and negative-energy spectra ${\mathcal E}_\pm(z)$
defined by Eq.~(\ref{energyl+-}) for $p_z=p_\perp=0$ to show
energy gap and level crossing.
From Eq.~(\ref{crosspoint+-}) we find now the
classical
turning points [instead of (\ref{scrossing})]
\begin{equation}
z_\pm = \frac{\ell}{2}\ln\left[2e^{({\mathcal E}\pm \sqrt{1+ \delta } )/\sigma}-1\right].
\label{hcrossing}
\end{equation}
For tunneling to take place,
the
 energy
${\mathcal E}$ has to satisfy
\begin{equation}
{\mathcal E} \le
 \sqrt{1+ \delta }
-\sigma\ln 2
,
\label{cregimeh}
\end{equation}
and $ \sigma $ must be larger than
$\sqrt{1+ \delta }\zeta $.
%
Expressing $z/\ell$ in terms of $\zeta$
as
\begin{equation}
z=z(p_\perp,\E;\zeta)=
\frac{\ell}{2}\ln\left[2e^{({\mathcal E}+ \zeta \sqrt{1+ \delta } )/\sigma}
-1\right],
\label{hcrossing2}
\end{equation}
so that $z_\pm=z(p_\perp,\E;\pm1)$,
we find
the electric field in the form
\begin{equation}
E(z)=E_0\left[1-{\frac{1}{2}}
e^{( \zeta\sqrt{1+ \delta }-\E)/ \sigma }\right]
\equiv
E(p_\perp,\E;\yy)
.
\label{hey}
\end{equation}
Inserting this into Eq.~(\ref{gf})
and expanding
$E_0/E(p_\perp,\E;\yy)$ in
 powers
we obtain
\begin{eqnarray}
G (p_\perp,{\mathcal E})=
1+\sum_{n=1}^\infty \frac{e^{-n\E/ \sigma }}{2^n} \frac{2}{\pi}
\int^{1}_{-1}
 d\zeta \sqrt{1-\zeta^2}\,e^{n \hat\zeta/ \sigma }
=
1+\sum_{n=1}^\infty e^{-n\E  / \sigma }
I_1(n\sqrt{1+ \delta } / \sigma ),
\end{eqnarray}
where $I_1(x)$
is a modified Bessel function.
Expanding
$I_1(n\sqrt{1+ \delta } / \sigma )$ in powers of $ \delta $:
\begin{eqnarray}
I_1(n\sqrt{1+ \delta } / \sigma )=
I_1(n / \sigma )+
(n/ 4\sigma )
[
I_0(n / \sigma )+
I_2(n / \sigma )] \delta +\dots~,
\end{eqnarray}
we identify
\begin{eqnarray}
G (0,{\mathcal E})&=& 1+
\sum_{n=1}^\infty e^{-n\E  / \sigma }
I_1(n/ \sigma ),  \\  ~
G (0,{\mathcal E})+
G _ \delta (0,{\mathcal E})
&=&1+      \frac{1}{2}
\sum_{n=1}^\infty e^{-n\E  / \sigma } [
(n/ \sigma )I_0(n/ \sigma )
-I_1(n/ \sigma )].
\end{eqnarray}
The integral over $\E$ in
Eq.~(\ref{pgwk1})
starts at $\E_<=1- \sigma \log 2$ where the integrand
rises from  0 to  $1$ as $\E$ exceeds a few units of $ \sigma $.
The derivative
of $e^{-\pi(E_c/E_0)G(0,\E)}$  drops from $1$  to
 $e^{-\pi(E_c/E_0)}$  over this  interval.
Hence the derivative
$\partial _\E
e^{-\pi(E_c/E_0)G(0,\E)}$ is  peaked
around some value $\bar \E$.
Thus we perform
the integral
$\int d\E e^{-\pi(E_c/E_0)G(0,\E)}$
by parts as
\begin{equation}
 \int d\E e^{-\pi(E_c/E_0)G(0,\E)}
 =
\E e^{-\pi(E_c/E_0)G(0,\E)}\Big|_{\E_<}^\infty-
 \int d\E \,\E\,\partial _\E e^{-\pi(E_c/E_0)G(0,\E)}.
 \label{hfc}
\end{equation}
The first term can be rewritten
with the help of
$d\E= eE_0dz$ as
$ e^{-\pi(E_c/E_0)}|eE_0|\ell/2$, thus giving rise to the
decay rate
(\ref{xy2}) in the volume $V_\perp \ell/2$ , and
the second term gives only a small correction to this. The second term in Eq.~(\ref{hfc}) shows that
the boundary effects reduce the pair-production rate compared with the pair-production rate (\ref{xy2})
in the constant field without any boundary.

\comment{
\begin{eqnarray}
A_1\equiv
 \sum_{n=0}^\infty\frac{n}{ \sigma }e^{-n\bar \E/ \sigma }I_1\left(\frac{n}{ \sigma }\right)=0,
\end{eqnarray}
where
it has the expansion
\begin{eqnarray}
 G (0,{\mathcal E})=A_0+\frac{1}{2}A_2 \E^2/2+\dots~,
\end{eqnarray}
with
\begin{equation}
A_2=  \sum_{n=0}^\infty e^{-n\bar \E/ \sigma }\frac{n^2}{ \sigma ^2}I_1\left(\frac{n}{ \sigma }\right),~~~~~
A_0=  \sum_{n=0}^\infty e^{-n\bar \E/ \sigma } I_1\left(\frac{n}{ \sigma }\right),
\end{equation}
Hence we find
\begin{equation}
\int d\E e^{-(E_c/E_0)\pi}
\end{equation}
}

\section{Tunneling into Bound States}

\begin{figure}[th]
\begin{center}
\begin{picture}(105.64,184.645)
\def\fsz{\footnotesize}
\def\ssz{\scriptsize}
\def\tsz{\tiny}
\def\dst{\displaystyle}\unitlength1mm
\put(-50,0){\includegraphics[width=6cm,clip]{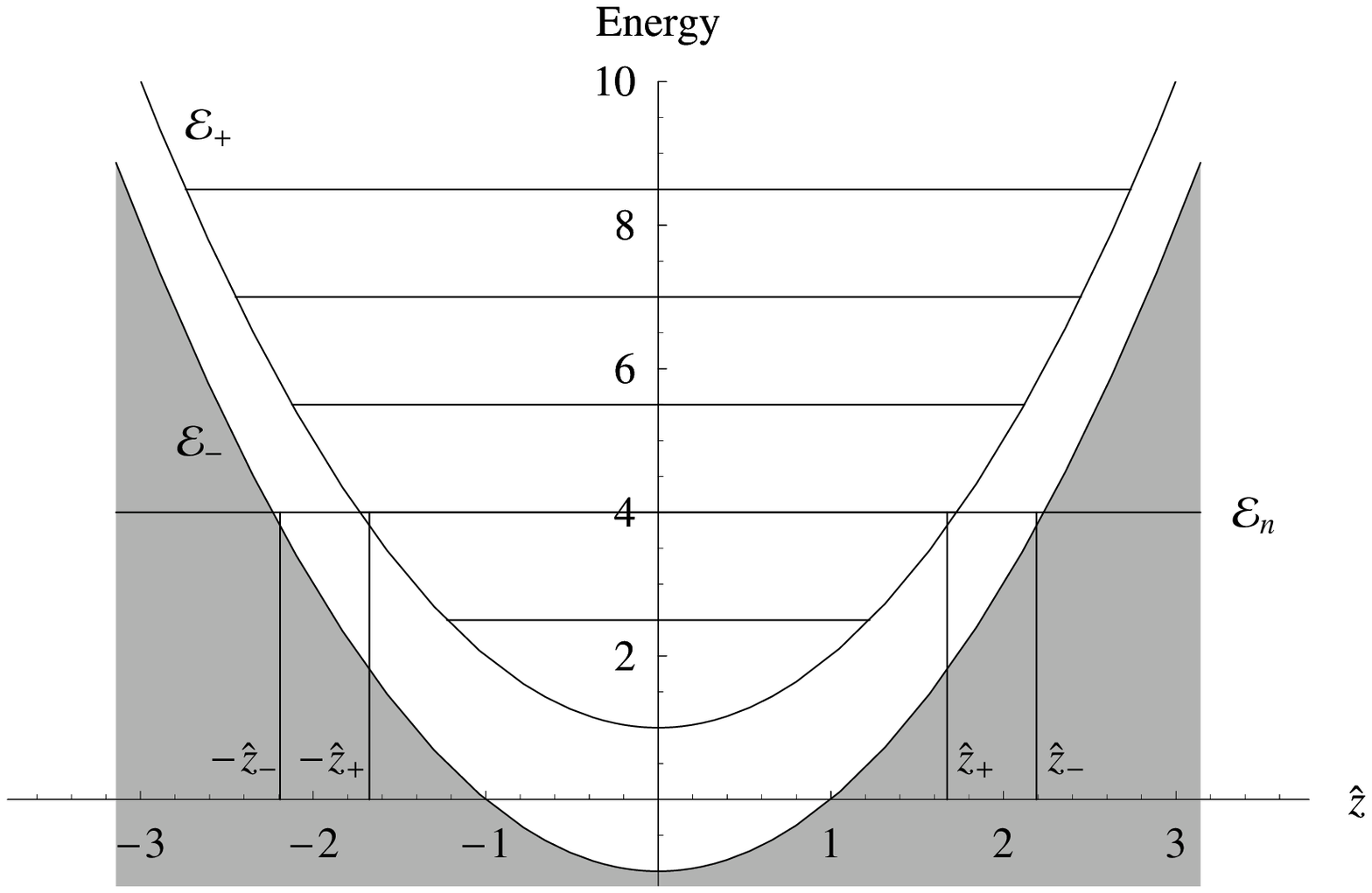}}
\put(30,0){\includegraphics[width=6cm,clip]{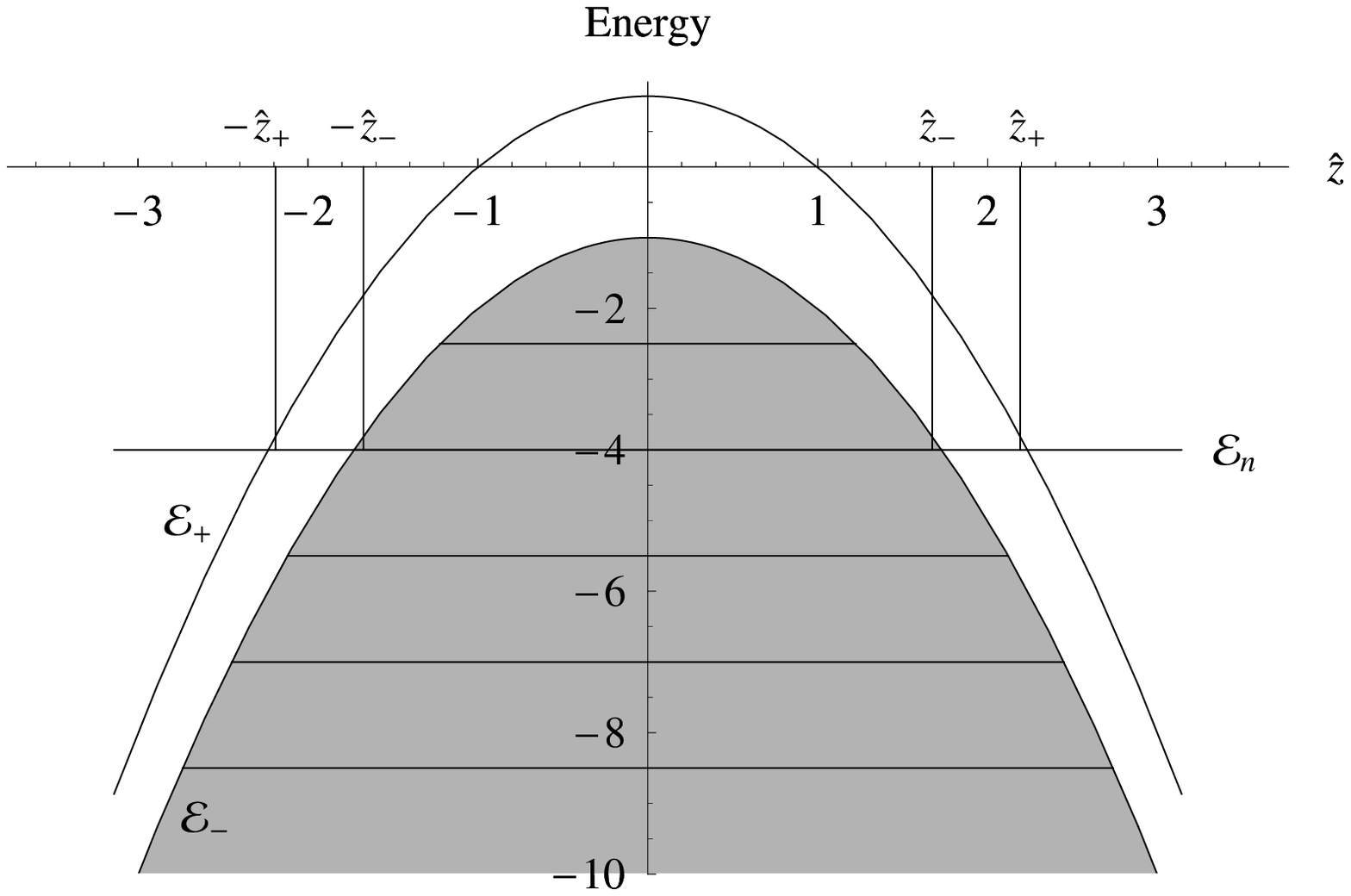}}
\end{picture}
\end{center}
\caption[]{
Positive- and negative-energy spectra ${\mathcal E}_\pm(z)$
of Eq.~(\ref{energyl+-}) for $p_z= p_\perp =0$
as a function of $\hat z\equiv z/\lambda_C$
for the linearly rising  electric field $E(z)$
with the harmonic potential (\ref{hav}). The
reduced field strengths are $ \epsilon =2$ (left figure) and
$ \epsilon =-2$ (right figure).
On the left, bound states
are filled and positrons escape to $z=\pm\infty$.
On the right,
bound electrons with negative energy
tunnel out of the well and escape with increasing energy to
$z=\pm\infty$.
}
\label{hamo}%
\end{figure}

We turn now to the case in which
instead of an outgoing wave as given (\ref{WKBout})
there is a bound state.
We consider a linearly rising electric field
whose potential is harmonic:
\begin{eqnarray}
\label{hav}
E(z)=E_0\left(\frac{z}{\lambda_C}\right),~~~\label{hafield}
V(z)= \frac{e  E_0\lambda _C}{2}\left(\frac{z}{\lambda_C}\right)^2.
\end{eqnarray}
It will be convenient to parametrize the field strength
$E_0$ in terms of a dimensionless
reduced electric field
 ${\epsilon}$ as
 $E_0= \epsilon \hbar c/e\lambda_C^2=\epsilon E_c$.
In
Fig.~\ref{hamo}
we have plotted the positive and negative-energy spectra ${\mathcal E}_\pm(z)$
defined by Eq.~(\ref{energyl+-}) for $p_z=p_\perp=0$ to show
energy gap and level crossing for $\epsilon >0$ (left) and $\epsilon<0$
(right).
If  $\epsilon$
is positive, Eq.~(\ref{crosspoint+-})
yields
for $z>0$,
\begin{equation}
z_\pm = \lambda_C \sqrt{ \frac{{2}}{\epsilon}}\left({\mathcal E}\mp \sqrt{1+ \delta } \right)^{1/2},
~~~~
z_+ < z_-,\label{hmcrossing}
\label{@LHS}\end{equation}
 and mirror-reflected
turning points for
$z<0$,
obtained by
exchanging $z_\pm \rightarrow -z_\pm $ in (\ref{@LHS}).
Negative-energy
electrons
tunnel
into the potential well $-z_+ < z < + z_+$, where $ {\mathcal E} \ge \E_+ $,
forming
bound states.
The associated  positrons run off to infinity.

\subsection{WKB transmission probability}

Due to the physical application to be discussed
in the next section, we shall study here only the
tunneling process
for $\epsilon > 0$
on the left-hand side
of Fig.~\ref{hamo}.
We consider the regime $z<0$
with the turning pints $-z_-<-z_+$.
The incident wave and flux for $z<-z_-$ pointing
in the positive $z$-direction are given by Eqs.~(\ref{WKBin})
and (\ref{influx}). The wave function
for  $-z_-<z<-z_+$ has the form
Eq.~(\ref{WKBt}) with the replacement $z_-\rightarrow -z_-$.
The transmitted
wave is now no
 longer  freely propagating as in (\ref{WKBout}), but a
describes a bound state
of a positive-energy electron: 
\begin{eqnarray}
\phi_{\E_n}(z)=\frac{\mathcal B}{(p_z)^{1/2}}\cos\left[ \frac{1}{\hbar}\int^z_{-z_+}p_zdz - \frac{\pi}{4}\right].
\label{WKBbo}
\end{eqnarray}
The Sommerfeld quantization condition
\begin{eqnarray}
\frac{1}{\hbar}\int_{-z_+}^{+z_+}p_z dz = \pi (n+\sfrac{1}{2}),~~~n=0,1,2,\dots~.
\label{quan}
\end{eqnarray}
fixes the
energies  $\E_n$. The connection
rules
for the wave functions (\ref{WKBt}) and (\ref{WKBbo})
at the turning point
 $-z_+$ determine
\begin{eqnarray}
{\mathcal B}=\sqrt{2}{\mathcal C}_+ e^{-i\pi n}\exp \left[ -\frac{1}{\hbar}\int_{-z_-}^{-z_+}\kappa_z dz\right].
\label{outbo}
\end{eqnarray}
Assuming the states $\phi_{\E_n}(z)$ to
 be initially unoccupied, the transmitted flux
to these states at the classical turning point $-z_+$ is
\begin{eqnarray}
\frac{\hbar}{m_e}\phi_{\E_n}(z)\partial_z\phi^*_{\E_n}(z)\Big|_{z\rightarrow -z_+}=\frac{|{\mathcal B}|^2}{(2m_e)}
=\frac{|{\mathcal C}_+|^2}{m_e}\exp \left[ -\frac{2}{\hbar}\int_{-z_-}^{-z_+}\kappa_z dz\right].
\label{fluxbo}
\end{eqnarray}
From Eqs.~(\ref{influx}), (\ref{fluxbo}), and (\ref{wdefine})
we then find the
WKB transmission probability for
 positrons
to fill these bound states
leaving a positron outside:
\begin{eqnarray}
W_{\rm WKB}(p_\perp,\E_n)=
\exp \left[ -\frac{2}{\hbar}\int_{-z_-}^{-z_+}\kappa_z dz\right]
,
\label{wdefine1}
\end{eqnarray}
which has the same form as Eq.~(\ref{tprobability1}).
The same result is obtained
once more
for $z>0$ with turning points $z_+<z_-$, which can be obtained
from (\ref{wdefine1}) by the mirror reflection
$-z_\pm\leftrightarrow z_\pm$.

\subsection{Energy spectrum of bound states}

From Eq.~(\ref{WKBr2}) for $p_z$ and Eq.~(\ref{hafield}) for the potential $V(z)$, we calculate
the eikonal (\ref{quan}) to determine the
energy spectrum $\E_n$ of bound states
\begin{eqnarray}
\frac{1}{\hbar}\int_{-z_+}^{+z_+}p_z dz &=& 2\frac{\epsilon}{\lambda_C^3}\int_{0}^{z_+}\left[(z^2-z_+^2)(z^2-z_-^2) \right]^{1/2} dz\nonumber\\
&=&\frac{2\epsilon z_+}{3\lambda_C^3}\left[(z_+^2+z_-^2)E(t)-(z_-^2-z_+^2)K(t)\right],
\label{quan1}
\end{eqnarray}
where $E(t)$, $K(t)$ are complete
elliptical integrals of the first and second kind, respectively, and $t\equiv z_+/z_-$.
The
Sommerfeld quantization rule (\ref{quan}) becomes
\begin{eqnarray}
\frac{8}{3}\left[\frac{2({\mathcal E}_n- \sqrt{1+ \delta })}{\epsilon} \right]^{1/2}\left[\E_n E(t_n)-(\sqrt{1+\delta})K(t_n)\right]=\pi (n+\frac{1}{2});\quad
t_n\equiv \left(\frac{{\mathcal E}_n- \sqrt{1+ \delta }}{{\mathcal E}_n + \sqrt{1+ \delta }} \right)^{1/2}.
\label{quan2}
\end{eqnarray}
For
any given
transverse momentum
$p_\perp= \sqrt{ \delta } $, this determines
the discrete energies
 $\E_n$.

\subsection{Rate of pair production}

By analogy
with Eqs.~(\ref{gxy0}) and (\ref{pgwk1}),
the transmission
probability (\ref{wdefine1}) must now be
integrated
over all
incident
particles with the flux (\ref{nflux})
to yield
the rate of pair production:
\begin{eqnarray}
\frac{\Gamma _{\rm WKB}}{V_\perp}
&=&
2\Ds \sum_n\frac{ \omega_n}{2\pi}
\int\frac{d^2{p}_\perp}{(2\pi\hbar)^2}
W_{\rm WKB}(p_\perp,{\mathcal E}_n),\label{gwkbpb0}\\
&\approx &
2\Ds
\frac{|eE_0 |}{4\pi^2\hbar c}
\sum_n \frac{\omega_n}{2\pi}
\frac{1}{
G(0,{\mathcal E}_n)
+G_ \delta (0,{\mathcal E}_n)
}
e^{-{\pi ( E_c/E_0)G(0,{\mathcal E}_n)}}.
\label{gwkbpb1}
\end{eqnarray}
In obtaining these
expressions
we
have
 used
the  energy conservation law to perform the integral over
$\E$. This receives
only contributions
for $\E=\E_n$
where  $\int d\E=\omega_n\hbar\equiv \E_n-\E_{n-1}$.
The factor $2$ accounts for the equal contributions from
the two regimes
$z>0$ and $z<0$.
The previous relation (\ref{@REL}) is
now replaced by
\begin{equation}
\omega_n\hbar = |eE(z_-^n)|\Delta z_-^n.
\end{equation}
Using Eq.~(\ref{y(x)}) and expressing $z/\lambda_C >0$ in terms of $\zeta$
as
\begin{equation}
z=z(p_\perp,\E_n;\zeta)=
\lambda_C \sqrt{\frac{{2}}{{\epsilon}}}\left({\mathcal E}_n-\zeta \sqrt{1+ \delta } \right)^{1/2},
\label{hacrossing2}
\end{equation}
we calculate $z_\pm=z(p_\perp,\E_n;\pm1)$,
and find
the electric field in the form
\begin{equation}
E(z)=E_0 \sqrt{\frac{{2}}{{\epsilon}}}\left({\mathcal E}_n-\zeta \sqrt{1+ \delta } \right)^{1/2}
\equiv
E(p_\perp,\E_n;\yy)
.
\label{hmey}
\end{equation}
Inserting this into Eq.~(\ref{gf}) we obtain
\begin{eqnarray}
G (p_\perp,{\mathcal E}_n)&=&\frac{2}{\pi}
 \sqrt{\frac{{\epsilon}}{{2}}}\int_{-1}^1d\zeta\frac{\sqrt{1-\zeta^2}}{[{\mathcal E}_n-\zeta \sqrt{1+ \delta }]^{1/2}},\nonumber\\
&=&
\frac{8}{3\pi}\sqrt{\frac{{\epsilon}}{{2}}}\frac{({\mathcal E}^\delta_n+1)^{1/2}}{(1+ \delta )^{1/4}}\left[(1-{\mathcal E}^\delta_n)
K (q^ \delta _n)+{\mathcal E}^\delta_nE(q^\delta_n)\right]
\label{hagf}
\end{eqnarray}
where ${\mathcal E}^\delta_n\equiv {\mathcal E}_n/(1+\delta)^{1/2}$
and $q^\delta_n=\sqrt{2/({\mathcal E}^\delta_n+1)}$.
Expanding
$G (p_\perp,{\mathcal E}_n)$ in powers of $\delta$ we find
the zeroth order term
\begin{eqnarray}
G (0,{\mathcal E}_n)&=&
\frac{8}{3\pi}\sqrt{\frac{{\epsilon}}{{2}}}({\mathcal E}_n+1)^{1/2}
\left[(1-{\mathcal E}_n)K(q_n)+{\mathcal E}_nE(q_n)\right]
\label{hagf0}
\end{eqnarray}
and
the derivative
\begin{eqnarray} \!\!\!\!\!
G_\delta (0,{\mathcal E}_n)&=&
\frac{\sqrt{\epsilon}}{3\pi}\frac{q_n}{{\mathcal E}_n}{q_n-1}\left[(4-5q_n
+{\mathcal E}_n(7-6q_n))E(q_n)+(1-{\mathcal E}_n-7{\mathcal E}_n^2)
(q_n-1)K(q_n)\right].
\label{hagf1}
\end{eqnarray}
where $q_n\equiv \sqrt{2/({\mathcal E}_n+1)}$.
\section{Coulomb electric field}

We now come to the physically interesting
system of a bare nucleus
of high charge $Z$
in the vacuum
which fills its empty bound states
by quantum tunneling
of
electrons
from
 the negative-energy continuum
around it.
The radial
Coulomb
field
and the
potential are given by
\begin{eqnarray}
eE(r)=\frac{ \hat \alpha \hbar c}{r^2},~~~
V(r)=-\frac{ \hat \alpha \hbar c}{r},
\label{cv}
\end{eqnarray}
where $\hat \alpha=Z\alpha$.
The Schr\"odinger equation
of this problem reads
\begin{eqnarray}
\left[ \left(\E+\frac{\hat \alpha \hbar c}{r} \right)^2+
c^2\hbar ^2\nabla^2-m_e^2c^4\right] \psi_\E({\bf x})=0.
\label{schr@}\end{eqnarray}
After factorizing
out spherical harmonics
$
\psi_\E({\bf x})=
R_{\E,l}(r)Y_{lm}(\hat{\bf x})$,
the radial wave functions satisfy
\begin{eqnarray}
\left\{ \E^2+
c\hbar \frac{2\E \hat \alpha }{r} +
c^2\hbar^2 \left[ \partial _r^2
-\frac{l(l+1)-\hat \alpha^2}{r^2}\right]
-m_e^2c^4
\right\}
 rR_{\E,l}(r)=0,
\label{@Semicl}\end{eqnarray}
where $l=0,1,2,3,\dots$
are the quantum numbers of
angular momentum.
The term in brackets cannot
 be treated semiclassically
since it contains
a factor $\hbar ^2$.
This destroys the possibility of
a systematic expansion
of the wave function
in
powers of $\hbar $.
In fact, the  $\hbar ^2/r^2$-potential
should be considered as being part
of the kinetic term, not of the potential.
In fact, both have the same scaling dimension.
There exists
a heuristic
way of accounting for this
due to Langer
\cite{LAN}.
He introduced a change of coordinates
to
$\xi\equiv \log r$,
so that
$\partial _r^2\, rR_{\E,l}(r)
=e^{-2\xi}(\partial ^2_\xi-\partial _\xi) r(\xi)R_{\E,l}(r(\xi))
=e^{-2\xi}(\partial ^2_\xi-{\scriptstyle\frac{1}{4}})e^{-\xi/2} r(\xi)R_{\E,l}(r(\xi))$,
which brings
(\ref{@Semicl})
to the form
\begin{eqnarray}
\bigg[r^2(\xi) (\E^2
-m_e^2c^4)+
2c\hbar {\E \hat \alpha }\,{r(\xi)} +
{c^2\hbar ^2}\left\{\partial _\xi^2
-[{l(l+1)+{\scriptstyle\frac{1}{4}}-\hat \alpha ^2}]
\right\}
\bigg]
  \sqrt{r(\xi)}R_{\E,l}(r(\xi))=0,
\label{@Semicl1}\end{eqnarray}
Now the $\hbar ^2/r^2$-part of the
kinetic term has become
a trivial constant
which
no longer influences the
semiclassical
treatment \cite{remofootnote}.

Equivalently, we can apply a
corrected semiclassical treatment directly
to Eq.~(\ref{@Semicl})
if
we add ${\scriptstyle\frac{1}{4}}$
to $l(l+1)$ in the numerator of the centrifugal barrier.
This
so-called
 {\em Langer correction\/}
is implemented
in (\ref{@Semicl})
 by replacing
$l(l+1)-\hat \alpha ^2$ by $ \lambda _l( \lambda_l +1)$, where
\begin{equation}
 \lambda _l
=\left[ \left(l + \sfrac{1}{2}
      \right)^2 - \hat \alpha ^2\right] ^{1/2 }-\frac{1}{2}
     = l-  \frac{\hat \alpha ^2 }{2 l + 1} +
    {\cal O}( \alpha ^4).
\label{lambda}\end{equation}
Note that Eq.~(\ref{@Semicl1})
has lost the singularity of Eq.~(\ref{@Semicl}) at the origin, a fact which is often considered to be the
motivation
for going to the variable $\xi$.
However, due to the
equal scale of $\hbar ^2/r^2$ and the gradient term $\hbar ^2\partial _r^2$,
this is {\em no\/}t the relevant property. The $\hbar ^2/4r^2$-correction is needed at
{\em any distance\/} to obtain the correct wave function at
the semiclassical level. In fact, this  wave function,
although being approximate,
turns out to
produces the {\em exact\/}
energy levels of the relativistic Coulomb system.

\subsubsection{Semiclassical quantization for point-like nucleus}

The semiclassical treatment
of Eq.~(\ref{@Semicl})
with $l$ replaced by $ \lambda _l$
starts out  with
the Langer-corrected energy
\begin{equation}
{\mathcal E}_\pm(p_r,l;r)=\pm\sqrt{(cp_r)^2+(\hbar c)^2
{ ( l+{\scriptstyle\frac{1}{2}})^2}/{r^2}+m^2_ec^4}+V(r).
\label{energyl+-C}
\end{equation}
We now impose the
Sommerfeld quantization rule
upon the
 eikonal:
\begin{equation}
S(\E)=\int_{r_{\rm i}}^{r_{\rm o}} dr\, p_r=\pi\hbar (n_r+{\scriptstyle\frac{1}{2}}),
\label{@EiKo}\end{equation}
where   $n_r=0,1,2,\dots$ is the radial quantum number, and
\begin{equation}
p_r= \frac{1}{c}\sqrt{
\E^2+2c\hbar \hat \alpha\E /r
-
c^2\hbar ^2 ( \lambda_l +{\scriptstyle\frac{1}{2}})^2/r^2-m_e^2c^4
}  ,
\label{prc}\end{equation}
whose zeros
yield  the turning points for each $l$:
\begin{eqnarray}
r_{{\rm o},{\rm i}}=\frac{c\hbar }{m_e^2c^4-\E^2}
\left[
\hat \alpha \E\pm \sqrt{ \hat \alpha^2\E^2+( \lambda_l +{\scriptstyle\frac{1}{2}})^2
(\E^2-m_e^2c^4)}\right] .
\label{@Inner}\end{eqnarray}
Consider first the energy regime $0<\E<m_ec^2$.
For
 $0<r_{\rm i}<r_{\rm o}$,
we
rewrite $p_r$ as
\begin{equation}
p_r=\frac{1}{c} \sqrt{m_e^2c^4-\E^2}\,\frac{1}{r}
 \sqrt{
\left(r_{\rm o}-{r}\right)
\left(r-{r_{\rm i}}\right)},
\label{pr1}\end{equation}
and perform the integral
in (\ref{@EiKo})
to find the eikonal
\begin{equation}
S(\E)= \frac{1}{c}
 \sqrt{m_e^2c^4-\E^2}\,\frac{\pi}{2}(r_{\rm o}+r_{\rm i}-2 \sqrt{r_{\rm o}r_{\rm i}})
=
\pi \hbar \left[
\frac{ \hat \alpha \E}{ \sqrt{m_e^2c^4-\E^2}}
-   ( \lambda _l+{\scriptstyle\frac{1}{2}})
\right] ,
\label{se1}\end{equation}
where\mn{check}
\begin{eqnarray}
r_{\rm o}r_{\rm i} & = & c^2\hbar^2\frac{( \lambda_l +{\scriptstyle\frac{1}{2}})^2}{m_e^2c^4-\E^2},\quad r_{\rm o}+r_{\rm i}= c\hbar\frac{2\hat \alpha \E}{m_e^2c^4-\E^2}.
\label{rori}
\end{eqnarray}
Inserting
$
S(\E)$ into the quantization condition (\ref{@EiKo}),
we obtain
 the
following equation for the
exact bound state energies
\begin{equation}
   \frac{\E_{nl}{} ^2 - m_e^2c^4}{2m_e^2c^4} = - \frac{\hat \alpha^2}{2}
     \frac{\E_{nl} {}^2}{m_e^2c^4} \frac{1}{(n_r+ \lambda_l +1)^2}.
\label{dell}\end{equation}
This is solved by
\begin{eqnarray}
  \E_{nl} & = & \pm\frac{m_ec^2}{ \sqrt{1 + (Z  \alpha) ^2/(n- \delta _l)^2} }
  \nonumber\\
    & = &\pm m_e c^2\left[  1- \frac{ \hat \alpha ^2}{2n^2} +
    \frac{3}{8} \frac{\hat \alpha^4}{n^4}-
    \frac{\hat \alpha ^4}{n^3 (2l + 1)}
      + {\cal O} ( \hat \alpha^6)\right] ,
\label{dell2}\end{eqnarray}
where
$n\equiv n_r+l+1$
is the principal quantum number of the atom, and
\begin{eqnarray}
     \delta _l & = &  \left(l +\sfrac{1}{2}\right) - \left(\lambda _l +\sfrac{1}{2}\right)
     =   \frac{\hat \alpha ^2 }{2 l + 1} +
    {\cal O}(\hat \alpha ^4).
\label{rel.287}
\end{eqnarray}
Although derived by
a semiclassical approximation,
these happen to be the exact values, due to the Langer correction
in (\ref{prc}).

For a Dirac particle, the energy-spectrum $\E_{nj}$ can be obtained
from a slight modification of
 Eq.~(\ref{dell2}):
\begin{eqnarray}
  \E_{nj} & = & \pm\frac{m_ec^2}{ \sqrt{1 + (Z  \alpha) ^2/(n- \delta _j)^2} }
\label{dell2D}\end{eqnarray}
where
$j=\sfrac{1}{2},\sfrac{3}{2},\cdot\cdot\cdot,n-\sfrac{1}{2}$
is
the total angular momentum.
For each $j$, there are two
degenerate states with orbital angular momentum
 $l=l_\pm=j\pm\sfrac{1}{2}$, for which we define
 $\lambda _{j_\pm}\equiv l_\pm- \delta _j$,
with
\begin{equation}
 \delta _j\equiv j+\sfrac{1}{2}- \sqrt{( j+\sfrac{1}{2})^2-\hat  \alpha ^2}
=\hat  \alpha ^2/(2j+1)+{\cal O}(\hat  \alpha ^4).
\label{dre1}\end{equation}
The radial quantum number
$n_r$ is related to
the others by
$
    n_r +  \lambda_{j_\pm}  + 1 = n_r + l_\pm + 1 -
\delta _l = n -  \delta _j.$

\subsubsection{Semiclassical quantization for finite-size nucleus}

According
 to  Eq.~(\ref{dre1}), the energy (\ref{dell2}) becomes imaginary
 for the $l=0$-state of spin-$0$ particle
when $\hat \alpha >1/2$, and
of a
spin-$\sfrac{1}{2}$ particle, when
$\hat \alpha >1$.
This signalizes the
crash of the electron
into the Coulomb potential of a
point-like positive nuclear charge.
The result is, however,
unphysical since nuclei always have a finite radius $r_{\rm n}$.
We account for this
by the
approximation
a
uniform distribution of charge inside the nucleus
\cite{z}, so that the potential is
\begin{equation}
V(r)=-\left\{\begin{array}{ll} \hbar c\,\lfrac{\hat \alpha}{ r}, &  r\ge r_{\rm n}, \\
 \hbar c\lfrac{\hat \alpha\,f \left(\lfrac{r}{ r_{\rm n}}\right)}{ r_{\rm n}}, &  r\le r_{\rm n},
 \end{array}\right.
\label{extpotential}
\end{equation}
where $f(x)\equiv (3-x^2)/2$.
Now the Sommerfeld quantization condition
for the eikonal (\ref{@EiKo}) reads
\begin{equation}
S(\E)=
S^{(1)}(\E)+S^{(2)}(\E)\equiv
\int_{\tilde r_0}^{r_{\rm n}} dr\,  p^{(1)}_r+\int_{r_{\rm n}}^{r_{\rm o}} dr\, p_r^{(2)}=\pi\hbar (n_r+{\scriptstyle\frac{1}{2}}),
\label{@EiKo1}
\end{equation}
with the radial momenta in the different regions:
%
\begin{eqnarray}
 p_r^{(1)} & = & \frac{1}{c}\sqrt{
\left[\E+\hbar c\frac{\hat \alpha}{ r_{\rm n}}f\left(\frac{r}{ r_{\rm n}}\right)\right]^2
-
c^2\hbar ^2\frac{(l+{\scriptstyle\frac{1}{2}})^2}{r^2}-m_e^2c^4
} ,\label{prvi}\\
 p_r^{(2)} &=&  \frac{1}{c}\sqrt{
\left[\E+\hbar c\frac{\hat \alpha}{ r}\right]^2
-
c^2\hbar ^2\frac{(l+{\scriptstyle\frac{1}{2}})^2}{r^2}-m_e^2c^4
}.
\label{prvo}
\end{eqnarray}
%

Let us  first calculate the eikonal $S^{(1)}(\E)$ inside
the nucleus. Expanding $
 p_r^{(1)}
$
in powers of $r/r_{\rm n}$  keeping terms up to the order
${\mathcal O}[(r/r_{\rm n})^2]$
we obtain
\begin{eqnarray}
 p_r^{(1)}&\approx &  \frac{1}{c}\left\{
\left(\E +c\hbar\frac{3\hat \alpha}{2r_{\rm n}}\right)^2
-m_e^2c^4-\frac{\hat \alpha}{r_{\rm n}}\left(
\E+c\hbar\frac{3\hat \alpha}{2r_{\rm n}}\right)\left(\frac{r}{r_{\rm n}}\right)^2
-c^2\hbar ^2\frac{(l+{\scriptstyle\frac{1}{2}})^2}{r^2}
\right\}^{1/2} \nonumber\\
&\approx &  \frac{1}{c}\left[
\left(\E +c\hbar\frac{3\hat \alpha}{2r_{\rm n}}\right)^2
-m_e^2c^4\right]^{1/2}\frac{1}{r}(r^2-\tilde r_0^2)^{1/2}
\label{prva}
\end{eqnarray}
where
\begin{equation}
\tilde r_0=
c\hbar \left[\frac{(l+{\scriptstyle\frac{1}{2}})^2}{(\E+c\hbar\frac{3\hat \alpha}{2r_{\rm n}})^2-m_e^2c^4}\right]^{1/2}
\approx \frac{2}{3\hat\alpha}(l+{\scriptstyle\frac{1}{2}})r_{\rm n}.
\label{zeroa}
\end{equation}
The approximation is good for  $r_{\rm n}/\lambda_C\ll 1$,
which is assured as long as
\begin{equation}
|\E|\ll \hbar c \frac{\hat\alpha}{r_{\rm n}},\quad {\rm and}\quad
\Lambda_l^2\equiv-(\lambda_l+{\scriptstyle\frac{1}{2}})^2=\hat\alpha^2-(l+{\scriptstyle\frac{1}{2}})^2>0.
\label{zeroa1}
\end{equation}
The second inequality determines
a maximum value of the angular-momentum $l$
for a given $\hat\alpha$.
Inserting the approximate value
(\ref{zeroa})  into (\ref{prva}), and the associated
$p_r^{(1)}$ into
(\ref{@EiKo1}),
we obtain
 the $r<r_{\rm n}$-part of the eikonal
\begin{eqnarray}
S^{(1)}(\E)& \approx & m_ec\left[(r^2-\tilde r_0^2)^{1/2}
+ r_{\rm n}(l+{\scriptstyle\frac{1}{2}})\arcsin\left(\frac{\tilde r_0}{r}\right)\right]^{r_{\rm n}}_{\tilde r_0}\nonumber\\
&=& \hbar \left(\frac{r_n}{\lambda_C}\right)\left\{\left[1-\left(\frac{2l+1}{3\hat \alpha}\right)^2\right]^{1/2}+ (l+{\scriptstyle\frac{1}{2}})\left[\arcsin\left(\frac{2l+1}{3\hat \alpha}\right)-\frac{\pi}{2}\right]\right\}.
\label{eiko3}
\end{eqnarray}

We now calculate the
 eikonal $S^{(2)}(\E)$ outside
the nucleus
in Eq.~(\ref{@EiKo1}).
We begin with the negative-energy regime
$-m_ec^2<\E<0$,
where  Eq.~(\ref{zeroa1}) is satisfied so that Eq.~(\ref{@Inner}) gives
$r_{\rm o}\gg r_{\rm n}$, and unphysical
turning points $r_{\rm i}<0$ and $|r_{\rm i}|>r_{\rm o}$.
Using Eq.~(\ref{pr1}) for $p_r^{(2)}(r)$ and
integrating it over
$r_{\rm n}<r<
r_{\rm o}$
we find
\begin{eqnarray}
&&\!\!\!\!\!\!\!\!\!\!\!\!\!\!\!\!\!\!S^{(2)}(\E)\equiv
\int_{r_{\rm n}}^{r_{\rm o}} dr\, p_r ^{(2)}=\frac{1}{c}\sqrt{m_e^2c^4-\E^2}\cdot\nonumber\\
&&\!\!\!\!\!\!\!\!\!\!\!\!\!\!\!\!\times \left[ R-(-r_{\rm o}r_{\rm i})^{1/2}\ln\frac{2(-r_{\rm o}r_{\rm i})+(r_{\rm o}\!+r_{\rm i})r
+2(-r_{\rm o}r_{\rm i})^{1/2}R}{r}-\frac{r_{\rm o}\!+r_{\rm i}}{2}\arcsin\frac{(r_{\rm o}\!+r_{\rm i})-2r}{(r_{\rm o}\!-r_{\rm i})}\right]^{r_{\rm o}}_{r_{\rm n}}\!,
\label{se2}
\end{eqnarray}
where $R(r)=\sqrt{
\left(r_{\rm o}-{r}\right)
\left(r-{r_{\rm i}}\right)}$,
and
\begin{eqnarray}
r_{\rm o}-r_{\rm i}=2c\hbar\frac{\sqrt{ \hat \alpha^2\E^2+( \lambda_l +{\scriptstyle\frac{1}{2}})^2
(\E^2-m_e^2c^4)}}{m_e^2c^4-\E^2}.
\label{rori-}
\end{eqnarray}
Under the conditions $r_{\rm n}/\lambda_C\ll 1$ and $r_{\rm n}/|r_{o,i}|\ll
 1$,
Eq.~(\ref{se2}) becomes
approximately
\begin{eqnarray}
S^{(2)}(\E)\approx \hbar\left[\frac{\hat \alpha\E}{\sqrt{m_e^2c^4-\E^2}}\left(\frac{\pi}{2}+\arcsin
\frac{\hat \alpha\E}{\chi_l}\right)-\Lambda_l\left( 1+\ln\frac{r_{\rm n}\chi_l}{c\hbar\Lambda_l^2}\right)\right],
\label{se2a}
\end{eqnarray}
where
\begin{eqnarray}
\chi_l\equiv \sqrt{\Lambda_l^2(m_e^2c^4-\E^2)+\hat \alpha^2\E^2}.
\label{chi}
\end{eqnarray}
For $r_{\rm n}\ll \lambda_C$,
the first eikonal part $S^{(1)}(\E)$
of Eq.~(\ref{eiko3}) is negligible compared with the second
part $S^{(2)}(\E)$ of Eq.~(\ref{se2a}).
Thus can simply do
the  calculation with a Coulomb potential
cut off at $r=r_{\rm n}$.
The charge distribution inside nucleus
is irrelevant here.

For negative energies $\E$ close to zero, we
approximate
\begin{eqnarray}
S^{(2)}(\E)\approx \hbar\left[\frac{1}{2\Lambda_l}\left(\hat\alpha^2+\Lambda_l^2\right)\left(\frac{\E}{m_ec^2}\right)^2
+\frac{\pi\hat\alpha}{2}\left(\frac{\E}{m_ec^2}\right)-\Lambda_l\left(1+\ln\frac{r_{\rm n}}{\lambda_C\Lambda_l}\right)\right],
\label{ese2a}
\end{eqnarray}
and the quantization rule (\ref{@EiKo1}) yields the energy levels
\begin{eqnarray}
\E_{nl}\approx -m_ec^2\frac{\pi\hat\alpha\Lambda_l+\sqrt{(\pi\hat\alpha\Lambda_l)^2+8(\hat\alpha^2+\Lambda_l^2)
\left[\pi\left(n_r+{\scriptstyle\frac{1}{2}}\right)+\Lambda_l(1+\ln\frac{r_{\rm n}}{\lambda_C\Lambda_l})\right]}}{2(\hat\alpha^2+\Lambda_l^2)}.
\label{esp1}
\end{eqnarray}
For negative energies $\E$ close to $-m_ec^2$,
expression (\ref{se2a})
simplifies to
\begin{eqnarray}
S^{(2)}(\E)\approx \hbar\left[\left(\frac{\Lambda_l}{2}-\frac{\Lambda_l^3}{6\hat\alpha^2}\right)\eta^2-\Lambda_l\left(2+\ln\frac{r_{\rm n}\hat\alpha}{\lambda_C\Lambda_l^2}\right)\right],
\label{ese3a}
\end{eqnarray}
where $\eta^2\equiv (m_e^2c^4-\E^2)/(m_ec^2)^2$, so that
the quantization rule (\ref{@EiKo1}) gives
\begin{eqnarray}
\E_{nl}\approx -m_ec^2\left[1-6\hat\alpha^2\frac{\pi\left(n_r+{\scriptstyle\frac{1}{2}}\right)+\Lambda_l\left(2+\ln\frac{r_{\rm n}\hat\alpha}{\lambda_C\Lambda_l^2}\right)}{3\hat\alpha^2\Lambda_l-\Lambda_l^3}\right]^{1/2}.
\label{esp2}
\end{eqnarray}
The critical value $\hat\alpha_c(l)$
can be obtained from Eq.~(\ref{@EiKo1}), and
from (\ref{ese3a}) for $\eta^2=0$,
\begin{eqnarray}
\Lambda_l^2=\hat\alpha_c^2-(l+{\scriptstyle\frac{1}{2}})^2\approx
\pi^2\left[\frac{n_r+{\scriptstyle\frac{1}{2}}}{2+\ln\frac{r_{\rm n}\hat\alpha}{\lambda_C\Lambda_l^2}}\right]^2.
\label{crita}
\end{eqnarray}
%

Consider now
the negative-energy regime below $-m_ec^2$
under the assumption (\ref{zeroa1}), so that
Eq.~(\ref{@Inner}) yields $r_{\rm o}\gg r_{\rm n}$,
 and unphysical zeros at $0<
r_{\rm o}<r_{\rm i}$. By
writing $p_r$ in Eq.~(\ref{prc}) by analogy with
in (\ref{pr1})
as
\begin{equation}
p_r=\frac{1}{c} \sqrt{\E^2-m_e^2c^4}\,\frac{1}{r}
 \sqrt{
\left(r_{\rm o}-{r}\right)
\left({r_{\rm i}-r}\right)},
\label{pr3}\end{equation}
we can perform the second integral in (\ref{@EiKo1})
and obtain
\begin{eqnarray}
&&\!\!\!\!\!\!\!\!\!\!\!\!\!\!\!\!\!S^{(2)}(\E)\equiv\int_{r_{\rm n}}^{r_{\rm o}} dr\, p_r =\frac{1}{c}\sqrt{\E^2-m_e^2c^4}\cdot\nonumber\\
&&\!\!\!\!\!\!\!\!\!\!\!\!\!\times \left[ \bar R-(r_{\rm o}r_{\rm i})^{1/2}\ln\frac{2(r_{\rm o}r_{\rm i})-(r_{\rm o}+r_{\rm i})r
+2(r_{\rm o}r_{\rm i})^{1/2}\bar R}{r}-\frac{r_{\rm o}+r_{\rm i}}{2}\ln[2\bar R+2r-(r_{\rm o}+r_{\rm i})]\right]^{r_{\rm o}}_{r_{\rm n}}\!,
\label{se2'}
\end{eqnarray}
where $\bar R(r)=\sqrt{
\left(r_{\rm o}-{r}\right)
\left({r_{\rm i}}-r\right)}$.
Under the conditions $r_{\rm n}/\lambda_C\ll 1$ and $r_{\rm n}/r_{o,i}\ll 1$,
we obtain
\begin{eqnarray}
S^{(2)}(\E)\approx \hbar \left[\frac{\hat \alpha |\E|}{\sqrt{\E^2-m_e^2c^4}}
\ln\left(\Lambda_l\frac{\sqrt{\E^2-m_e^2c^4}}{\hat \alpha |\E|}-1\right)
-\Lambda_l\left( 1+\ln\frac{r_{\rm n}\chi_l}{c\hbar\Lambda_l^2}\right)\right].
\label{se2'a}
\end{eqnarray}
Assuming $\hbar c\,\hat \alpha/r_{\rm n}\gg |\E| \gg m_ec^2$, we expand $S^{(2)}(\E)$ in terms of $m_ec^2/|\E|$ and the
leading order is
\begin{eqnarray}
S^{(2)}(\E)\approx \hbar \Lambda_l\left[\ln \frac{m_ec^2}{|\E|} +\frac{\hat\alpha}{\Lambda_l} \ln\left(\frac{\Lambda_l}{\hat\alpha}-1\right)
-1-\ln\frac{r_{\rm n}}{\lambda_C\Lambda_l^2}-\ln(\hat\alpha^2-\Lambda_l^2)\right],
\label{se2'ap}
\end{eqnarray}
and the quantization rule (\ref{@EiKo1}) yields energy-levels:
\begin{eqnarray}
\E_{nl}\approx -m_ec^2\frac{\lambda_C}{r_{\rm n}}
\left[\frac{\Lambda_l^2\left(\frac{\Lambda_l}{\hat\alpha}-1\right)^{\frac{\hat\alpha}{\Lambda_l}}}{\hat\alpha^2-\Lambda_l^2}\right]
\exp\left[-\frac{\pi}{\Lambda_l}(n_r+{\scriptstyle\frac{1}{2}})-1\right].
\label{se3'ap}
\end{eqnarray}
As for a spin-$\sfrac{1}{2}$ particle, the critical value $\hat\alpha_c(j)$ and energy-spectra $\E_{nj}$ are obtained from
the critical value (\ref{crita}) and energy-spectra
(\ref{esp2}) and (\ref{se3'ap}) by
the following replacement:
\begin{eqnarray}
\Lambda_l^2=\hat\alpha^2 -\left(l+\sfrac{1}{2}\right)^2~\Rightarrow~
\Lambda_j^2=\hat\alpha^2 -\left(j+\sfrac{1}{2}\right)^2.
\label{dre2}
\end{eqnarray}
As before, for each $j$
there are two
degenerate levels of orbit angular momentum $l=l_\pm=j\pm \sfrac{1}{2}$.

\subsubsection{WKB transmission probability}

\begin{figure}[th]
\begin{center}
\begin{picture}(105.64,144.645)
\def\fsz{\footnotesize}
\def\ssz{\scriptsize}
\def\tsz{\tiny}
\def\dst{\displaystyle}\unitlength1mm
\put(-20,0){\includegraphics[width=8cm,clip]{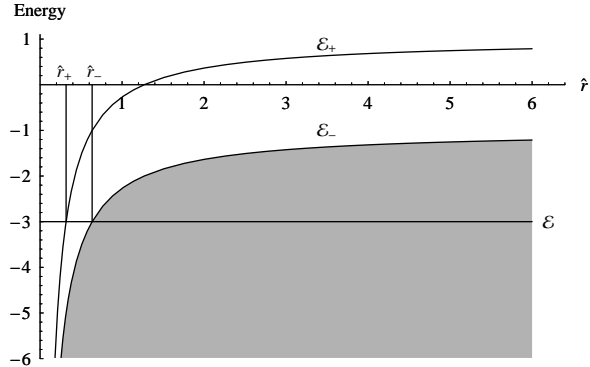}}
\end{picture}
\end{center}
\caption{For a radial Coulomb field,
the positive and negative-energy spectra ${\mathcal E}_\pm(r)$
of Eq.~(\ref{energyl+-})
are plotted
as a function of $\hat r=r/\lambda_C$. They are found by solving
condition $p_r=0$ at $l=0$
and  $\hat \alpha=1.27$ ($Z=174$).
}
\label{coulomb1}%
\end{figure}

The Schr\"odinger equation Eq.~(\ref{@Semicl}) can be written as
\begin{eqnarray}
\left[c^2\hbar ^2\frac{d^2}{dr^2}+c^2p^2_r(r)\right] rR_{\E,l}(r)=0,
\label{KGr}\end{eqnarray}
where
\begin{eqnarray}
c^2p_r^2=\left[\E-V(r) \right]^2-c^2\hbar^2\frac{(l+{\scriptstyle\frac{1}{2}})^2}{r^2}-m_e^2c^4,
\label{pr}
\end{eqnarray}
which
looks like
the Klein-Gordon equation (\ref{KG}) in one dimension
along the $r$-axis.
As before, we assume the nuclear
radius $r_{\rm n}$ to be much smaller
than the Compton wavelength  $\ll \lambda_C$,
Outside the nucleus,
$V(r)$ is the Coulomb potential.
From th
condition
$p_r=0$
we calculate
the classical turning points
$r_{\rm n}\ll
r_+<
r_-$,
by analogy with Eq.~(\ref{crosspoint+-}).
For a give energy $\E < -m_ec^2$, we have three regions
 (see Fig.~\ref{coulomb1}):
\begin{enumerate}
\item[(i)]
$r_-<r$
and $\E < \E_-$ ,
\item[(ii)]
$r_+<r<
r_-$
and $\E_- < \E <\E_+$,
\item[(iii)] $r<r_+$ and $\E > \E_+$.
\end{enumerate}
Starting (i)
where $p_r^2>0$, Eq.~(\ref{KGr})  has two independent solutions
corresponding to $p_r<0$ (incident wave)
and $p_r>0$ (reflected wave),
\begin{eqnarray}
R_{\E,l}=\frac{\mathcal C_+^r}{(p_r)^{1/2}}\frac{1}{r}\exp\left[
 \frac{i}{\hbar}\int^rp_rdr\right]  +
\frac{\mathcal C_-^r}{(p_r)^{1/2}}\frac{1}{r}\exp
\left[ -\frac{i}{\hbar}\int^rp_rdr\right] ,
\label{WKBinr}
\end{eqnarray}
where ${\mathcal C_\pm^r}=e^{\pm i\pi/4}{\mathcal C^r}/2$.
The corresponding solution
in the region (ii)
is given by
\begin{eqnarray}
\frac{{\mathcal C^r}}{2(\kappa_r)^{1/2}}\frac{1}{r}\exp\left[ -\frac{1}{\hbar}\int^{r_-}_{r}\kappa_rdr\right]+
\frac{\bar {\mathcal C^r}}{2(\kappa_r)^{1/2}}\frac{1}{r}\exp\left[ \frac{1}{\hbar}\int^{r_-}_{r}\kappa_rdr\right],
\label{WKBtr}
\end{eqnarray}
where $\kappa_r=-ip_r$ and $p_r^2<0$.
For a purely incident wave,
only the second term in $R_{\E^-,l}$
 is present,
with an incident flux density at $r=r_-$:
\begin{eqnarray}
j_r\equiv \frac{\hbar}{2m_ei}\left[\phi_r^*\hat\partial_r\phi_r - (\hat\partial_r\phi_r^*)\phi_r\right]
=-\frac{p_r}{m_e}\phi_r^*\phi_r=-\frac{|{\mathcal C}_-^r|^2 }{m_er^2_-},
\label{influxr}
\end{eqnarray}
where $\phi_r=rR_{\E,l}^-$ and $\hat\partial_r=(1/r)(\partial/\partial r)r$.
The superscript - indicates the inward flux.

We consider bound states of energy $\E<- m_ec^2$ confined within
the region $(\tilde r_0,r_+)$, where $\tilde r_0<r_{\rm n}$ is another classical
turning point of positive-energy branch $\E_+$
inside the nucleus, in addition to $r_+$.
Its value is obtained
from
Eq.~(\ref{pr}) for $p_r=0$:
\begin{eqnarray}
\E-V(\tilde r_0)=+\sqrt{c^2\hbar^2\frac{(l+\sfrac{1}{2})^2}{\tilde r^2_0}+m_e^2c^4}.
\label{pr0}
\end{eqnarray}
By analogy with Eq.~(\ref{WKBbo}), the semi-classical wave function of bound states is,
\begin{eqnarray}
R_{\E,l}^{\rm bs}=\frac{\mathcal B^r}{p_r^{1/2}}\frac{1}{r}\cos\left[ \frac{1}{\hbar}\int^r_{\tilde r_0}p_rdr - \frac{\pi}{4}\right],
\label{WKBbor}
\end{eqnarray}
satisfying
Sommerfeld's quantization rule
\begin{eqnarray}
\frac{1}{\hbar}\int^{r_+}_{\tilde r_0}p_r dr = \pi \left(n_r+\sfrac{1}{2}\right),\quad n_r=0,1,2,\cdot\cdot\cdot ,
\label{quanr}
\end{eqnarray}
For each $n_r=0,1,1\dots~$,
the solutions
yield the discrete energies $\E_{nl}$
with the principal quantum number
$n\equiv n_r+l+1$, where the angular momentum $l$
can take the values $l=0,\dots,n$.
The continuity of wave functions (\ref{WKBtr}) and (\ref{WKBbor})
at the classical point $r_+$ leads to
\begin{eqnarray}
{\mathcal B^r}=\sqrt{2}{\mathcal C}_-^r e^{-i\pi n_r}\exp \left[ -\frac{1}{\hbar}\int_{r_+}^{r_-}\kappa_r dr\right].
\label{outbobr}
\end{eqnarray}
Assuming the bound states $R_{\E,l}^{\rm bs}$ be unoccupied, the transmitted flux to these states at the classical
turning point $r_+$ is
\begin{eqnarray}
\frac{\hbar}{m_e}R_{\E,l}^{\rm bs}\,\hat\partial_r\,[R_{\E,l}^{\rm bs}]^*\bigg|_{r\rightarrow r_+}
=-\frac{|{\mathcal B}^r|^2}{2m_er_+^2}=-\frac{|{\mathcal C}_-^r|^2}{m_er_+^2}\exp \left[ -\frac{2}{\hbar}\int_{r_+}^{r_-}\kappa_r dr\right].
\label{fluxbor}
\end{eqnarray}
From Eqs.~(\ref{influxr}), (\ref{fluxbor}), and (\ref{wdefine})
we obtain
the transmission probability
of an
electron
to tunnel into the bound state:
\begin{eqnarray}
W_{\rm WKB}(\E,l)=\frac{r_-^2}{r_+^2}\exp \left[ -\frac{2}{\hbar}\int_{r_+}^{r_-}\kappa_r dr\right].
\label{wdefine1r}
\end{eqnarray}
Normalizing the incident flux density (\ref{influxr}) at $r_-$ we have
\begin{eqnarray}
j_r
\equiv \sum _l j_r^l= \sum _l
D_sv_r(r_-)
\frac{(2l+1)}{4\pi r_-^2}\int \frac{dp_r }{2\pi\hbar},
\label{influxr1}
\end{eqnarray}
and the rate of pair-production
in the state with angular momentum $l$ becomes:
\begin{eqnarray}
\frac{\Gamma_{\rm WKB}(\E,l)}{V_\perp}&=&W_{\rm WKB}(\E,l)j_r^l\nonumber\\
&=&D_s v_r(r_-)
\frac{(2l+1)}{4\pi r_+^2}\int \frac{dp_r}{2\pi\hbar}
\exp \left[ -\frac{2}{\hbar}\int_{r_+}^{r_-}\kappa_r dr\right]\nonumber\\
&=&D_s
 \frac{(2l+1)}{4\pi r_+^2}\int \frac{d\E}{2\pi\hbar}
\exp \left[ -\frac{2}{\hbar}\int_{r_+}^{r_-}\kappa_r dr\right],
\label{wdefine2r}
\end{eqnarray}
where $v_r(r_-)=\partial\E/\partial p_r|_{r=r_-}$.
This is evaluated further
in the same way as
Eqs.~(\ref{gwkbpb0}) and (\ref{gwkbpb1}):
 the
integral over $\E$
has only contributions
from the bound state energies $\E=\E_{nl}$,
so that
$\int d\E$ is equal to
$\omega_{nl}\hbar$,
where $\omega_{nl}=\E_{nl}/\hbar$ is the frequency of the bound state.
As a result, the sum over all Eq.~(\ref{wdefine2r}) takes the form
\begin{eqnarray}
\frac{\Gamma_{\rm WKB}}{V_\perp}&=&
D_s \sum_{nl}\frac{(2l+1)}{4\pi r_+^2}\frac{\omega_{nl}}{2\pi}
\exp \left[ -\frac{2}{\hbar}\int_{r_+}^{r_-}\kappa_r dr\right].
\label{wdefine3r}
\end{eqnarray}

\subsubsection{Sauter exponential factor in Coulomb potential}

For brevity we use natural units
where
$r$ is measured in units of
the  Compton wavelengths
$\lambda _C=\hbar /m_ec$,
and
 $\E_{nl}$ in units of $m_ec^2$.
By setting
 $p_r=0$ in
Eq.~(\ref{prc}),
we obtain the analog of Eq.~(\ref{y(x)}):
\begin{eqnarray}
\frac{1 }{r(l,\E_{nl};\zeta)}= \frac{{\mathcal E}_{nl}\hat\alpha+
\zeta\sqrt{(l+{\scriptstyle\frac{1}{2}})^2({\mathcal E}_{nl}^2-\zeta^2)+\hat\alpha^2}}
{(l+{\scriptstyle\frac{1}{2}})^2\zeta^2-\hat\alpha^2}.
\label{cyx}
\end{eqnarray}
The role of $p_\perp^2$
is now played by $(l+{\scriptstyle\frac{1}{2}})^2/r^2$.
The classical
turning point $r_-({\mathcal E}_{nl},l)$
where the negative-energy states can tunnel to the
positive ones
at $r_+({\mathcal E}_{nl},l)$
is
obtained from Eq.~(\ref{cyx}) by inserting $\zeta=\pm 1$.
We identify $r_+\equiv r_{\rm o}$ in previous sections. Since $r_- > r_+$,
the electrons move inwards to neutralize the Coulomb potential,
the positrons are pushed outwards to infinity.
At the highest energy ${\mathcal E}_{nl}=-1$
where tunneling can occur,
we find $r_-=+\infty$ and $r_+=[\hat\alpha^2-
(l+{\scriptstyle\frac{1}{2}})^2]/2\hat\alpha$.

For each angular momentum
$l$ and energy-level $\E_{nl}$, we obtain the Sauter exponential factor
from the analogs of
Eqs.~(\ref{wwkbp}) and (\ref{gf}):
\begin{eqnarray}
\exp \left[ -\frac{2}{\hbar}\int_{r_+}^{r_-}\kappa_r dr\right] &= & \exp\left\{ -\pi\frac{E_c}{E_0}
G(l,\E_{nl})\right\},
\label{cwkbp}
\end{eqnarray}
and
  \begin{eqnarray}
G(l,\E_{nl})\equiv \frac{2}{\pi}\int^{1}_{-1} d\yy
\left[1+\frac{(  l  +\frac{1}{2})^2}
{r^2(l,\E_{nl};\zeta)}\right]
\frac{\sqrt{1-\yy^2}}{E(l,\E_{nl};\yy)/E_0}
.
\label{lgf}
 \end{eqnarray}
Note that the bracket
in the integral is the analog of the prefactor
$
 \left[1+{(c{p}_\perp) ^2}/{m_e^2c^4}\right]$ in (\ref{wwkbp}).
Here it
can no longer be taken out of the integral.
Instead, there is now another simplification.
Since ${E}= \hat\alpha \hbar c/r^2$,
the function $G(l,\E_{nl})$ becomes
  \begin{eqnarray}
G(l,\E_{nl})\equiv \frac{2}{\pi}\int^{1}_{-1} d\yy\sqrt{1-\yy^2}
\left[r^2(l,\E_{nl};\zeta)+(  l  +\sfrac{1}{2})^2\right],
\label{lgf1}
 \end{eqnarray}
and with $r(l,\E;\zeta)$ from
(\ref{cyx}):
  \begin{eqnarray} \!\!\!  \!
\!\!\!\!G(l,\E_{nl})\!\equiv \!\frac{2}{\pi}\int^{1}_{-1} d\yy\sqrt{1\!-\!\yy^2}
\left\{\!\frac{[(  l  \!+\!\sfrac{1}{2})^2
\zeta^2\!-\!\hat\alpha^2]^2[(\E_{nl}\hat\alpha)^2\!+\!\zeta^2
((  l \! +\!\sfrac{1}{2})^2(\E_{nl}^2\!-\!\zeta^2)+\hat\alpha^2)]}
{[(\E_{nl}\hat\alpha)^2\!-\!\zeta^2((  l  \!+\!\sfrac{1}{2})^2(\E_{nl}^2\!-\!\zeta^2)\!+\!\hat\alpha^2)]^2}+(  l \! +\!\sfrac{1}{2})^2\!\right\}
\!.
\label{lgf2}
 \end{eqnarray}
The result of this integral is
surprisingly simple
remembering that
$r_\pm$ are given
by $r_\pm(\E_{nl};l)$
of
Eq.~(\ref{cyx}) for $\zeta=\pm 1$:
  \begin{eqnarray}
G(l,\E_{nl})=2\left(  l  +\sfrac{1}{2}\right)^2\left(\E_{nl}^2-
|\E_{nl}|\sqrt{\E_{nl}^2-1}\right)+2\hat\alpha^2\left(\frac{\sqrt{\E_{nl}^2-1}}{|\E_{nl}|}-1\right),
\label{lgf3}
\end{eqnarray}
valid for $\E_{nl}<-1$ and $\hat\alpha \ge \hat\alpha{}_c(l)$.
Inserting this into the Sauter exponential factor (\ref{cwkbp}), and this further into
 (\ref{wdefine3r}),
we obtain the rate of filling the empty
bound state levels around the
Coulomb potential of the nucleus.
The sum over $l$
extends
to
the largest value
permitted by Eq.~(\ref{zeroa1}) and $\hat\alpha{}_c$
of (\ref{crita}).

The result for  a spin-$\sfrac{1}{2}$ particle
can be obtained
by replacing $\E_{nl}\Rightarrow \E_{nj}$ in
Eqs.~(\ref{cyx}) and (\ref{lgf3}), and
\begin{eqnarray}
\left(  l  +\sfrac{1}{2}\right)^2\Rightarrow \left(  j  +\sfrac{1}{2}\right)\left(  j \pm 1 +\sfrac{1}{2}\right),
 \label{dre3}
 \end{eqnarray}
where $j\pm 1$ respectively corresponds the state of parity $(-)^{l_\pm}$
with orbital angular momentum $l_\pm=j\pm \sfrac{1}{2}$.

\comment{
\mn{I do not think anyone has obtained this result for any $l$.}
If the negative-energy
states outside are empty, and the bound state
$\E_{nl}$
 is occupied,
it can tunnel towards the outside with a rate
\mn{connect this text logically with previous}
\begin{eqnarray}
\Gamma _{\rm WKB}(l)=\omega_{\rm att}(\E)
W_{\rm WKB}(l,{\mathcal E})
\Big|_{\E\rightarrow \E_{nl}},
 \end{eqnarray}
where
$\omega_{\rm att}(\E)
$
is the
attempt frequency
\begin{equation}  \!\!\!\!\!\!\!\!
\omega_{\rm att}(\E)=
\frac{1}{2t(\E)
} \equiv \frac{1}{2\partial S(\E)/\partial \E}
\end{equation}
in the bound state.
}

\section{Summary and remarks}

By studying the
process of
electron-positron pair production
from the vacuum
by
  a nonuniform electric field
as a
quantum tunneling phenomenon
 we have derived
in semiclassical approximation
the general rate formulas
(\ref{pgxy3}) and (\ref{pgwk1}).
They consist of
a Boltzmann-like tunneling
exponential,
 and a pre-exponential factor, and are applicable
to any system where the field strength points mainly
in
one direction and varies only along this direction.
The formulas
require the evaluation of the
functions
$G(0,\E)$ of Eq.~(\ref{gfhbar})
and
$G_\delta(0,\E)$ of Eq.~(\ref{ghbar}).
This has been done for several different
field configurations.

For  electrons arriving
by tunneling in a bound state
of a harmonic
electric potential, the general expressions are
 given by Eqs.~(\ref{gwkbpb1}) and (\ref{wdefine3r}) as functions
of the frequencies $\omega_n$ of the bound states.
The
discrete energy levels $\E_n$
at fixed $p_\perp$
are found from the
eikonal and Sommerfeld quantization. For fermions,
the expression for pair-production rate should
be multiplied by the Pauli-blocking factor for the rate
which makes it zero
if the bound state $\E_n$ is
occupied.

In the Coulomb electric field $E(r)=eZ/r^2$ of a nucleus with finite radius $r_{\rm n}\ll \lambda_C$,
we have given first the
semiclassical energy-levels $\E_{nl}$
in Eqs.~(\ref{dell2}), (\ref{esp1}), (\ref{esp2}), (\ref{se3'ap}), and
the formulas for the associated
pair production rate in (\ref{wdefine3r}), (\ref{cwkbp}), (\ref{lgf})
 for $\E_{nl}\le -m_ec^2$. The critical value
$\hat  \alpha _c\equiv Z_c\alpha$
is found
from
 Eq.~(\ref{crita})
 as a function
of the principal quantum number $n$ and the angular momentum $l$, which agree with the one found
in Refs.~\cite{z,z10}
the $n=1$-state
$1S_{\sfrac{1}{2}}$.

The number of energy-levels $\E_{nl}\le -m_ec^2$
that are able to accommodate electrons produced
from the vacuum is limited,
and pair production ceases when all these levels are fully occupied
even if the electric field is
overcritical.



\begin{thebibliography}{99}
\bibitem{remark}

By setting $\Ds$ equal to $1$ one can obtain the tunneling result also for
spin-$0$ particles although the Dirac picture is no longer applicable.
\bibitem{REMA2}
This treatment is analogous to that of the translational degree of freedom
in instanton calculations in Section 17.3.1
of \cite{PI} [see in particular Eq.~(17.112)].
\bibitem{remofootnote}
This term will be examined within the original coordinate
as a purely relativistic effect in a coming paper.
\end{thebibliography}
\end{document}